\documentclass[
 reprint,
 amsmath,amssymb,
 superscriptaddress,
]{revtex4-1}

\usepackage[all,defaultlines=2]{nowidow} 

\usepackage[section]{placeins} 
\usepackage{color} 
\usepackage{cancel} 
\usepackage{lineno} 
\usepackage[utf8]{inputenc} 
\usepackage{natbib} 
\usepackage{csquotes} 
\usepackage[caption = false]{subfig} 
\usepackage{graphicx}
\usepackage{dcolumn}
\usepackage{bm}
\usepackage{textcomp}
\usepackage{upgreek}
\usepackage{hyperref} 
\hypersetup{colorlinks=true, linkcolor=blue, filecolor=magenta, urlcolor=blue}  
\usepackage{float}

\begin{document}

\title{Heat emitting damage in skin: a thermal pathway for mechanical algesia}

\author{Tom Vincent-Dospital}
\email{tom.vincent-dospital@fys.uio.no}
\affiliation{SFF Porelab, The Njord Centre, Department of physics, University of Oslo, Norway}

\author{Renaud Toussaint}
\affiliation{Université de Strasbourg, CNRS, Institut Terre \& Environnement de Strasbourg, UMR 7063, France}
\affiliation{SFF Porelab, The Njord Centre, Department of physics, University of Oslo, Norway}

\author{Knut J\o rgen M\aa l\o y}
\affiliation{SFF Porelab, The Njord Centre, Department of physics, University of Oslo, Norway}

 
\begin{abstract}

         \textbf{Abstract}: Mechanical pain (or mechanical algesia) can both be a vital mechanism warning us for dangers or an undesired medical symptom important to mitigate. Thus, a comprehensive understanding of the different mechanisms responsible for this type of pain is paramount. In this work, we study the tearing of porcine skin in front of an infrared camera, and show that mechanical injuries in biological tissues can generate enough heat to stimulate the neural network. In particular, we report local temperature elevations of up to $24$ degrees Celsius around fast cutaneous ruptures, which shall exceed the threshold of the neural nociceptors usually involved in thermal pain. Slower fractures exhibit lower temperature elevations, and we characterise such dependency to the damaging rate. Overall, we bring experimental evidence of a novel - thermal - pathway for direct mechanical algesia. In addition, the implications of this pathway are discussed for mechanical hyperalgesia, in which a role of the cutaneous thermal sensors has priorly been suspected. We also show that thermal dissipation shall actually account for a significant portion of the total skin's fracture energy, making temperature monitoring an efficient way to detect biological damages.
         
\end{abstract}
 
\maketitle

\section*{Introduction}

\begin{figure*}
  \includegraphics[width=1\linewidth]{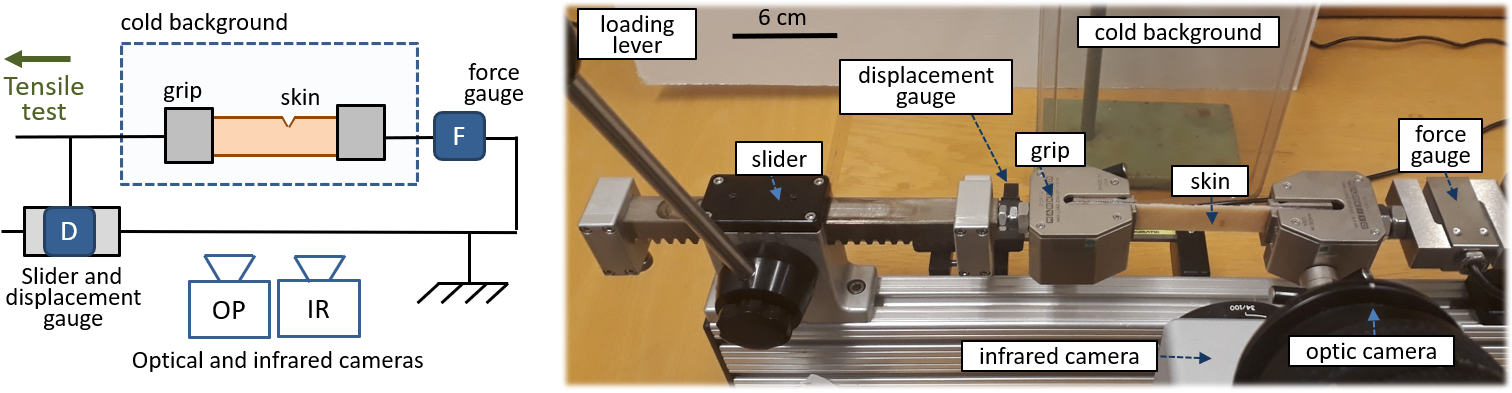}
  \caption{Schematic (left) and picture (right) of the experimental set-up. Porcine skin is teared in front of an optical and an infrared camera, while both the applied force and displacement are measured. In the real set-up (e.g., by contrast to what is suggested by the schematic) the optical camera is placed above the infrared one.}
  \label{fig:setup}
\end{figure*}

The toughness of matter is often characterised by its energy release rate\,\cite{Griffith1921}, that is, by the amount of mechanical energy which is dissipated by the rupture of a surface unit of a given solid matrix. Among other dissipation mechanisms (e.g.,\,\cite{rice_surface,crackwaves}), running cracks tend to emit some heat, a phenomenon which has long been reported and studied by the fracture physics community (e.g.,\,\cite{Irwin1957, RiceLevy, Fuller1975, Bouchaud2012}). In various materials, such as PMMA (polymethyl methacrylate)\,\cite{TVD2}, acrylic adhesives\,\cite{TVD2} or paper\,\cite{ToussaintSoft}, such heat actually accounts for a significant portion (i.e., from $10$ to almost $100$\%) of the total energy release rate, that is, it accounts for a significant portion of these materials' strength. More than a simple energetic loss, the heat dissipation have been suspected to lead to potentially paramount secondary effects, which are still actively debated. These effects notably include the brittleness of matter\,\cite{Marshall_1974, carbonePersson, ThermalRunaway, ToussaintSoft, TVD1} or the instability of some seismic faults (e.g.,\,\cite{HeatWeak, pressur2005, SulemCarbo}) through various thermal weakening phenomena.\\
Recently, we have also suggested a theory\,\cite{TVDpain1} that the damage-induced heat in biological tissues could be responsible for a degree of mechanical pain in the human body. Indeed, should the heat dissipation be large enough, it may be detected by the thermo-sensitive nociceptors of sensory neurons, and, in particular, some of the so-called TRP proteins. Action potentials (i.e., electrochemical signals) may thus be triggered in the nervous system. TRPs (standing for Transient Receptor Potential cation channels - e.g.,\,\cite{PainTRP}) are proteins expressed in many cells of the cutaneous tissue\,\cite{skinTRP}, and, in particular, at the surface of sensory neurons. Their ability to gate ions through cells' membranes\,\cite{voltage_gating} is temperature dependent, and different TRP types react to different cold or warm temperature ranges. As we are here interested in the detection of hot anomalies around running fractures, let us now focus on the TRPs that have been reported to be heat sensitive in skin. In our previous theoretical work\,\cite{TVDpain1}, the role of TRPV3 and TRPV1 was considered. The former (TRPV3) is sensitive to temperatures in the normal biological range (i.e., $30$ to $40^\circ$C)\,\cite{TRVP3_grad2}, making it, likely, responsible in part for the feeling of warmth. The latter (TRPV1) starts to activate at more painful temperatures above about $43^\circ$C\,\cite{TRPV1_julius}. Additionally to TRPV3, other nociceptors, TRPV4\,\cite{TRPV4_temp} and TRPM2\,\cite{TRPM2_temp}, are also responsive to subtle temperature changes in the normal physiological range, and, in addition to TRPV1, TRPM3 has also been evidenced to detect higher painful temperatures\,\cite{TRPM3_temp}. Completing the ranges of these various heat sensors, TRPV2, activates at the most noxious temperatures above $52^\circ$C, although it has been suggested that this particular sensor has little role in mechanical or thermal pain\,\cite{non_pain_TRPV2}. Generally, the role of TRPs is commonly admitted in thermal sensing, and some also react to specific chemicals, for instance contained in `hot' pepper or `cool' mint (e.g.,\,\cite{PainTRP}). The role of TRPs in the feeling of mechanical pain, which is here of main interest, has also been previously suspected. Thermal and mechanical pains were shown to be coupled in human subjects\,\cite{coupled_pain}, with a threshold to feel mechanical pain that decreases at a higher ambient temperature. Incidentally, cooling is sometimes used for the anesthesia of cutaneous and non-cutaneous tissues prior to medical mechanical injections (e.g.,\,\cite{cool_anesthesia,cool_anesthesia2}). In rodents, the drug-induced inhibition and activation of TRPV1 and TRPV3 has also proven to, respectively, reduce or increase mechanical hyperalgesia\,\cite{coupled_pain, TRPV1_mechano,TRPV1_mechano2,TRPV3_mechano} (i.e., the decreased threshold to feel mechanical pain after a first stimulus). It should however be noted that the neuroscience community is not unanimous when it comes to the involvement of TRPV1 in mechanical hyperalgesia (e.g.\,\cite{non_hyper_TRPV1}). Finally, the involvement of mammalian TRPV4 in the direct mechanosensation of living organisms has also been shown\,\cite{TRPV4_mechano}.\\
In this work, we experimentally show that the rupture of skin can generate heat anomalies that are in the sensing ranges of the mentioned TRP proteins, on time and space scales that are similar to those of these nociceptors sensibility\,\cite{neurite_density, TRP_response2, TRP_response}. We thus confirm the relevance of the proposed new thermal pathway for mechanical algesia. We tear pork skin samples, assumed to be a reasonably good model for human skin (e.g.,\,\cite{PigHuman, PigHuman2, skin_mod_examp}), in front of an infrared camera and report temperature elevations, over hundreds of milliseconds, of a few degrees to tens of degrees depending on the skin samples and the damaging rate. With a normal skin temperature of about $35^\circ$C\,\cite{skin_surf_temp,skin_temp1}, such thermal anomalies shall indeed open the TRP channels, and we here discuss both direct algesia and hyperalgesia scenarii. We characterise the relationship between damage velocity and local temperature elevation, suggesting that a minimal fracture velocity of about $1$\,cm\,s\textsuperscript{-1} may be needed for strong thermo-mechanical pain to actually be at play. We also provide the energy release rate of our samples, $\sim 135\,$kJ\,m\textsuperscript{-2} in average, and, with two different methods, we give a coarse estimation of the portion this energy release rate ($\sim3$\% to $50$\%) which actually transforms into heat, when fractures progress in skin. We thus show that heat dissipation is responsible for a non negligible part of the cutaneous strength, actually making temperature monitoring an efficient way to detect mechanical damages in biological tissues.

\section{Methods}

\subsection{Experimental set-up}

Let us start by describing the experimental set-up. Most of it is a standard mechanical test bench, a schematic and a picture of which is shown in Fig.\,\ref{fig:setup}. Porcine skin samples are placed in this horizontal test bench, where they are held at their extremities by two self-tightening wedge grips. This type of grips holds stronger and stronger as the sample they clamp is brought into tension. Mechanical tensile tests are performed on the skin up to its full rupture. To that extent, one of the wedge grips can be displaced along a uniaxial slider with the help of a manual lever. The other grip is fixed. It is attached to a force sensor (Mark-10\textsuperscript{\textregistered} MR01-300) that allows force measurements up to $1$\,N accuracy. The displacement of the moving grip is monitored with a digital scale (Mitutoyo\textsuperscript{\textregistered} 572-311-10) having a $30\,\upmu$m precision. The accuracies of the force and displacement sensors are satisfactory in regard to the typical maximal force ($\sim 500$\,N) and maximal displacement ($\sim 2$\,cm) of our typical experimental realisations. An optical camera also records both the set-up and the deforming skin sample. A mono-channel infrared camera (Flir\textsuperscript{\textregistered} SC3000), measuring the radiation intensity in the $800$ to $900$\,nm bandwidth and equipped with a macroscopic lens, is placed in front of the skin. This infrared camera, which is sensitive to temperature changes of fractions of degree, monitors the sample as it is loaded up to rupture. It is the main measurement of these experiments, and we will further discuss, in section\,\ref{sec:resol}, how the time and space resolution of this particular camera is adequate to the present study. Finally, behind the skin sample, a cold plastic plate, just out of a freezer, acts as a cold background for the infrared images. This ensures that any measured elevation of temperature (i.e., the signal of interest) does arise from the sample itself, and later simplifies the images analysis. This cold plate is not in direct contact with the skin and is placed - arbitrarily - $20$\,cm away from it, so it does not affect the sample's temperature by direct heat conduction.\\
Experimental set-ups, similar to the one we have here described (i.e., set-ups which monitor a fracture with an infrared camera), have regularly measured significant temperature elevations in the rupture of various materials (e.g.,\,\cite{Fuller1975, Bouchaud2012, ToussaintSoft}). However, such set-ups were never used, to our knowledge, for the heat characterisation of rupturing biological tissues.

\subsection{Skin samples and model limitations\label{sec:sample}}

\begin{figure}
  \includegraphics[width=1\linewidth]{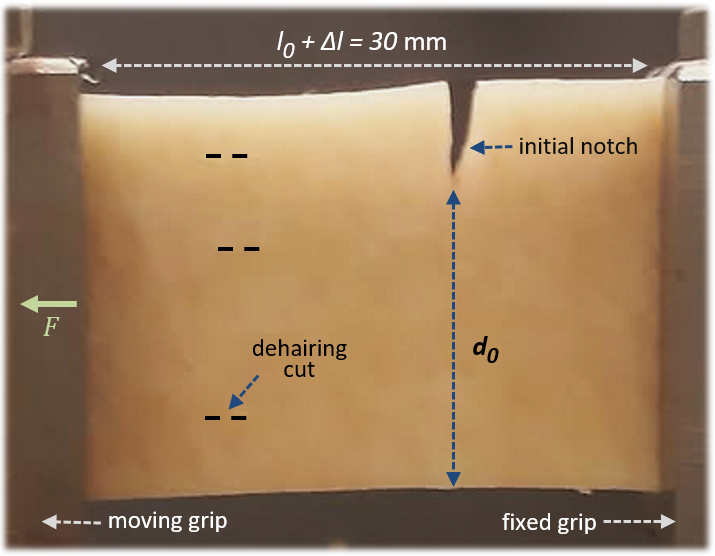}
  \caption{A porcine skin sample installed in the set-up, with the epidermis facing the camera. A little stretch, $\sim5\%$, is here applied by the force $F$, so that the nominal sample length between the grips is $l_0=28.5\,$mm and the stretch is $\Delta l=1.5\,$mm. The initial unbroken width is denoted $d_0$. The dashed lines illustrate some barely visible (shallow) cuts on the epidermis, from the dehairing process. The picture's point of view corresponds approximately to that of the infrared camera during the experiments.}
  \label{fig:sample}
\end{figure}
Porcine skin was here chosen as a model for human skin, as they both display similar structural, thermal and mechanical characteristics\,\cite{PigHuman, PigHuman2, skin_speheat2, skin_thermal}. Such comparison holds particularly well when comparing human skin to the cutaneous tissues of other mammals. The skin that we tested was acquired at a local butchery and originates from the flank or the upper leg of various pig specimens. It is a standard product sold by this butchery, and, thus, no pig was harmed specifically for the need of our experiment. The studied skin included the epidermis (that is, the surface layer of the skin), the dermis and a thin layer of subcutaneous fat (hypodermis). The total skin thickness varied between $1.6$ and $2.7$\,mm, and was measured on each sample with a caliper.
\\The rupture of seven skin specimens (denoted I to VII in this manuscript) was studied, in order to grasp some of the diversity in behaviours of such a biological material. 
With a scalpel, each skin sample was carved out to be rectangular with width $2.2$\,cm and length $10$ to $15$\,cm. The length of the samples simply allowed enough skin to be grabbed by the test bench's wedge grips (grabbing $3.5$\,cm on each side), to insure the samples stability during the experiments. The force, applied to tear the skin, was applied parallel to this direction. The width of the samples matched both approximately the grips' width and the frame's size of the infrared camera. Figure\,\ref{fig:sample} shows an optical picture of a skin sample installed in the set-up. To ensure that the skin rupture does occur in the frame of the infrared camera, a notch of length $3$ to $6\,$mm was initially cut perpendicularly to the sample's length, to control the initiation of the fracture.\\
We do not present in vivo experiments, for obvious ethical and practical reasons, and we should report that the tested skin went through two initial processes before being acquired that could have slightly alter its mechanical properties. The first of this processes is dehairing (e.g.,\,\cite{meat_processing}), which sparsely left some cuts on the epidermis (i.e., the shallowest part of our samples). 
This wear, if significant, should only have weakened the skin, as a perfectly intact skin would be overall tougher and, thus, would likely dissipate more heat upon rupture. In this regard, the results we here present are thus conservative. This being stated, we chose skin samples where the dehairing damage was the less pronounced. 
The second potentially altering process is freezing, which can cause micro-damages in soft biological tissues (e.g.,\,\cite{freezing}). The skin samples were indeed bought frozen and were unfrozen in a ambient temperature water bath during half an hour before running the rupture tests. After this bath, the samples' surfaces were only gently dried with a paper towel. In the case of porcine skin, it was shown\,\cite{PigHuman2} that the fracture energy of the epidermis significantly increases with the freezing/unfreezing process, although by less than an order of magnitude. Conveniently, in the same study (focusing on micro-needle insertions in the epidermis rather than on the full tearing of skin that we here consider) fresh human skin was actually shown to be as tough as unfrozen porcine one\,\cite{PigHuman2} (i.e., fresh human skin tend to be slightly stronger than fresh pork skin).\\
Overall, it should be stated that there is no perfect model for fresh human skin, and we here assume that the various physical parameters, measured on our butchery-acquired samples, are representative of our own cutaneous tissue.

\begin{figure*}
  \includegraphics[width=1\linewidth]{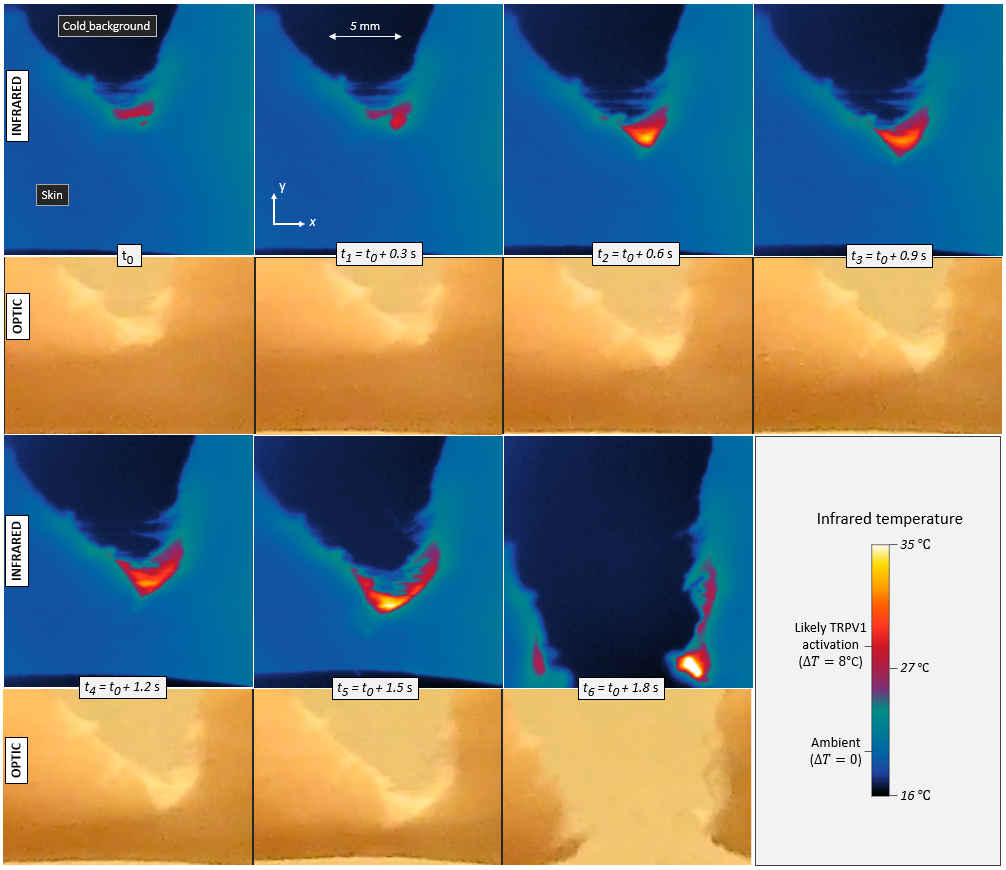}
  \caption{Infrared temperature maps and corresponding optical frames during the propagation of a tensile crack in porcine skin (skin specimen V). Seven different times are displayed with a constant increment in between. The maximum temperature elevation is $\Delta T=8^\circ$C in the first frame ($t_0$) and up to $\Delta T=17^\circ$C in the last one ($t_6$), Contrarily to the infrared camera, the optical camera monitored the whole set-up and was not equipped with a macroscopic lens, hence the noticeable difference in image resolution. It also recorded the skin with a slightly different (non-orthogonal) line of sight, as per Fig.\,\ref{fig:setup}.}
  \label{fig:result}
\end{figure*}

\subsection{Experimental versus biological resolutions\label{sec:resol}}

The displacement's digital measurement was outputted with a $5$\,Hz rate. The optical camera, recording with a $50$\,fps (frames per second) rate, hence allowed a faster but less accurate monitoring of the displacement. Depending on the experiment, the force measurement was outputted with a $10$ to $50$\,Hz rate. Overall, because the total loading time of the samples up to rupture was about ten seconds, these rates of measurement were satisfactory for the general characterisation of our mechanical tests. The infrared camera - i.e., the core measurement of our experiments - recorded with a $50$\,fps frame rate. It had a resolution of $240\,\times\,320$\,pixels, with a pixel size of $80$ to $90\,\upmu$m depending on the exact camera position and focus of each experimental realisation. This actual pixel size, for given realisations, was calibrated by capturing a hot metallic rod of known diameter, in the same plane as the skin sample on which the camera was priorly focused.\\
To understand how significant would any heat anomaly be in terms of algesia (pain), it is important to compare the time and space resolutions of our (infrared) thermal measurements to the time and space sensibility of the human nervous system. Based on microscopic observations\,\cite{neurite_density} of the density $D_n$ of neurites in the human epidermis (i.e., the body extensions of neurons), $D_n\sim2000$\,mm\textsuperscript{-2}, one can broadly estimate\,\cite{TVDpain1} the nominal distance between two of these neurites to be about $1/\sqrt{D_n}\sim20\,\upmu$m. At the surface of these neurites, the typical response time of some of the TRPs nociceptors to temperature jumps has also been measured\,\cite{TRP_response2,TRP_response}, with patch clamp experiments, to range from a few milliseconds to a few tens of milliseconds. Thus, our experimental resolution in space ($\sim85\,\upmu$m) and time ($1/50$\, Hz $=20\,$ms) should be rather close, yet slightly coarser, to that of neural sensing in the human skin. Any significant temperature change (i.e., as per the TRPs sensitivity) recorded by our set-up should then be able to initiate action potential in a live biological system. Oppositely, our infrared camera could miss some of the most accurate details available to the actual neural system on the smallest spatial and temporal scales. 

\section{Results}

\subsection{Temperature profiles}

Infrared videos, for all experiments, are available to the reader as supplementary materials. Figure\,\ref{fig:result} shows the maps of measured temperature $T$ at seven successive times for skin specimen V. As expected, some heat is emitted by the fracture as it progresses through the cutaneous tissue.\\
In order to convert the infrared signal to temperatures, skin was assumed to be a black body, that is, to follow Planck's law (\,e.g.,\,\cite{blackbody}) and have an emissivity close to $1$. In practice, the emissivity of porcine skin may range\,\cite{pig_emissivity} between $0.94$ and $1$. Varying this parameter, it was found that our reported elevations of temperature $\Delta T$ may hold an error of less than $10$\%, and this uncertainty will be irrelevant to our final conclusions.\\
In Fig.\,\ref{fig:result}, one can for instance observe temperature elevations up to $15^\circ$C ($\pm 1^\circ$C). This magnitude is significant with regard to general mammal biology and, more specifically, with regard  to the sensitivity of given neuronal thermal sensors. Indeed, assuming a normal inner temperature of about $35^\circ$C\,\cite{skin_surf_temp,skin_temp1} (i.e., for live subjects), elevations of $\sim8^\circ$C and more should be enough to trigger TRPV1 (and likely TRPM3), and elevations above about $17^\circ$C should trigger TRPV2\,\cite{PainTRP,skinTRP}. Additionally, fast thermal elevations of a few degrees Celsius could also excite TRPV3, TRPV4 or TRPM2. It was shown\,\cite{TRVP3_grad2}, in particular, that TRPV3 produces a higher bio-current intensity for faster rises in temperature, and, for temperature elevations of a few degrees, we here measured heating rates of up to $200^\circ$C s\textsuperscript{-1}. The thermal anomalies typically spread over a few millimetres across and along the crack trajectory, so that many neural receptors (at the surface of about $10^{4}$ neurites) could likely sense it. As priorly discussed, the typical spacing between two neurites should indeed be in the tens of micrometer range\,\cite{neurite_density}. Appendix\,\ref{app:all} presents some temperature profiles measured in the rupture of the six other skin specimens. Similar orders of magnitude are observed, although a variety of patterns shows in the temperature maps. The maximal temperature elevation $\Delta T_\text{max}$, which was recorded during each test, is indicated in table\,\ref{table:summary}. In all occurrences, it exceeds the TRPV1 threshold and sometimes exceeds that of TRPV2.

\subsection{Mean stress at rupture and elastic modulus}

\begin{figure}[b]
  \includegraphics[width=1\linewidth]{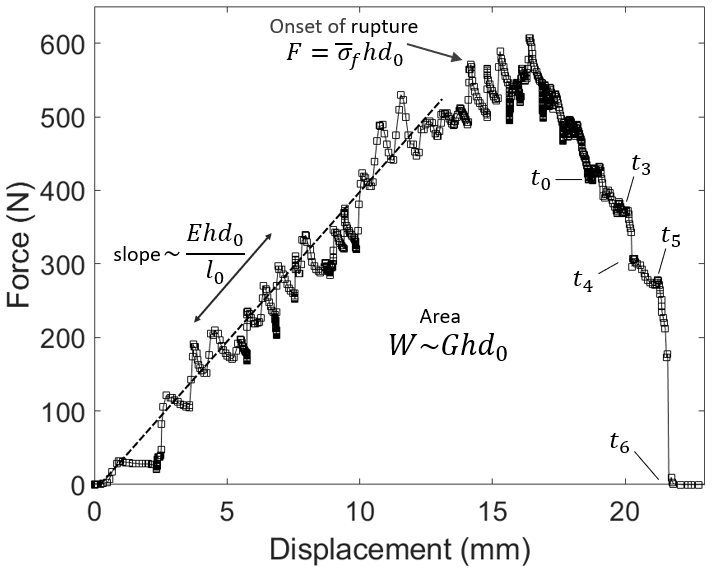}
  \caption{Force versus displacement plot for the experiment shown in Fig.\,\ref{fig:result} (skin specimen V). Labels $t_0$ to $t_6$ refer to the times of the frames in this other figure. The relatively linear relation at small displacements allows to invert for the skin's Young modulus $E$. The area below the plot is the total mechanical work $W_\text{tot}$ provided to the sample. The mean stress at the onset of rupture is called $\overline{\sigma}_\text{f}$.}
  \label{fig:force_disp}
\end{figure}

As we performed relatively standard tensile tests (the most exotic feature being to monitor them with an infrared camera), we here also provide some of the mechanical constants of our skin samples. From Fig.\,\ref{fig:force_disp}, showing for specimen V the measured force $F$ versus displacement $\Delta l$ plot, one can, in particular, estimate a mean stress $\overline{\sigma}_\text{f}$ at rupture by computing $\overline{\sigma}_\text{f}=F/(hd_0)$ at the onset of the fracture. Here, $h$ is the thickness of the sample and $d_0$ is its initial unbroken width. Note that this stress value does not account for any stress concentration, so that the actual stress shall reach, locally, higher values, around the initial crack tip or at the scale of the skin's collagen fibers (i.e., the main dry constituent of the dermis). The strength of our samples ranged from about $9$ to $19$\,MPa, which is rather logically comparable to the strength of individual collagen fibers (e.g.,\,\cite{single_fiber}). From the force versus displacement plots, an elastic (Young) modulus $E$ of the skin samples can also be estimated, from the approximately constant ratio $E = F l_0/(hd_0\Delta l)$ that holds as the sample is loaded elastically. In this expression, $l_0$ is the initial sample length between the two grips. We derived $E$ in the range of $20$ to $110$\,MPa. For each skin specimen, appendix\,\ref{app:all} shows the force and displacement plots and table\,\ref{table:summary} summarises the samples initial geometry (e.g.,\,$l_0$, $d_0$ and $h$), the computed values for the mean stress at rupture $\overline{\sigma}_\text{f}$ and the cutaneous Young modulus\,$E$.\\
Although not the core interest of the present study, it is satisfying that both the values of $E$ and $\overline{\sigma}_\text{f}$ that we report are compatible with other studies of ex vivo (e.g., post-mortem) skin samples (e.g.,\,\cite{other_young}). It is however to be noted that significantly lower elastic modulus, in the range of $1\,$Mpa, were also reported for human and porcine skins (e.g.,\,\cite{PigHuman2}). Extensively discussed in the literature (e.g., see\,\cite{other_young}), the causes for such variability may include the inherent spread in the properties of biological materials, the differences in testing methods (in particular for ex vivo versus in vivo samples), or anisotropy in the skin's structure. It may also lie in the dependence of the elastic modulus with the tissue's deformation rate\,\cite{Persson_skin}, as skin is a viscoelastic, rather than a purely elastic, material.

\subsection{Energy release rate}

\begin{table*}
\begin{tabular}{c|ccc|ccccc}
\begin{tabular}[c]{@{}c@{}}Skin\\ specimen\end{tabular} & \multicolumn{1}{c}{\begin{tabular}[c]{@{}c@{}}$h$\\ \,\,\,\,\,(mm)\,\,\,\,\,\end{tabular}} &
\multicolumn{1}{c}{\begin{tabular}[c]{@{}c@{}}$l_0$\\ \,\,\,\,\,(mm)\,\,\,\,\,\end{tabular}} & \multicolumn{1}{c|}{\begin{tabular}[c]{@{}c@{}}$d_0$\\ \,\,\,\,\,(mm)\,\,\,\,\,\end{tabular}} & \multicolumn{1}{c}{\begin{tabular}[c]{@{}c@{}}$\overline{\sigma}_\text{f}$\\ \,\,\,\,\,(MPa)\,\,\,\,\,\end{tabular}} & \multicolumn{1}{c}{\begin{tabular}[c]{@{}c@{}}$E$\\ \,\,\,\,\,(MPa)\,\,\,\,\,\end{tabular}} & \multicolumn{1}{c}{\begin{tabular}[c]{@{}c@{}}$\Delta T_\text{max}$\\ \,\,\,\,\,($^\circ$C)\,\,\,\,\,\end{tabular}} & \multicolumn{1}{c}{\begin{tabular}[c]{@{}c@{}}$G$\\ \,\,\,\,\,(kJ m\textsuperscript{-2})\,\,\,\,\,\end{tabular}} & \multicolumn{1}{c}{\begin{tabular}[c]{@{}c@{}}$\phi$\\ \,\,\,\,\,(-)\,\,\,\,\,\end{tabular}} \\ \hline
I & $2.0$ & $65$ & $18$ & $15\pm2$ & $110\pm20$ & $24\pm1.5$ & $130\pm20$ & $\sim30$\%\\
II & $2.0$ & $71$ & $16$ & $13\pm2$ & $31\pm6$ & $22\pm1.5$ & $160\pm20$ & $\sim5$\%\\
III & $1.8$ & $29$ & $18$ & $10\pm2$ & $35\pm6$ & $22\pm1.5$ & $80\pm10$ & $\sim30$\%\\
IV & $1.6$ & $32$ & $17$ & $9\pm2$ & $48\pm8$ & $19\pm1$ & $80\pm10$ & $\sim15$\%\\
V & $2.7$ & $25$ & $17$ & $12\pm2$ & $20\pm4$ & $17\pm1$ & $150\pm20$ & $\sim30$\%\\
VI & $2.0$ & $32$ & $14$ & $19\pm3$ & $33\pm6$ & $12\pm1$ & $210\pm30$ & $\sim15$\%\\
VII & $2.4$ & $42$ & $19$ & $10\pm2$ & $35\pm6$ & $16\pm1$ & $135\pm20$ & $\sim50$\%
\end{tabular}
\caption{Summary of various physical quantities estimated on each skin specimen.
$\overline{\sigma}_\text{f}$ is the mean stress at rupture, $E$ is the skin's Young modulus, $\Delta T_\text{max}$ the maximal temperature elevation recorded during a test, $G$ the mean energy release rate and $\phi$ the mean thermal efficiency in the dissipation of $G$. As explained in the text, $\phi$ should not be interpreted beyond its order of magnitude.
For reference, the initial samples geometry (i.e., as per Fig.\,\ref{fig:sample}) is also provided, with accuracy $\pm0.15$\,mm for $h$ and $\pm0.5$\,mm for $l_0$ and $d_0$.}
\label{table:summary}
\end{table*}

The rise in skin temperature, which we are here mainly interested in, accounts for a portion of the dissipated energy, as the rupture progresses. The total mechanical work that was provided during a tensile test is given by
\begin{equation}
    W_\text{tot} = \int_{\Delta l=0}^{+\infty} F \,\mathrm{d}\Delta l,
    \label{eq:work}
\end{equation}
that is, the area below the measured force versus displacement curve (i.e., see Fig.\,\ref{fig:force_disp}). By definition, the mean energy release rate of skin $G$ can then be derived as
\begin{equation}
    G \sim \frac{W_\text{tot}}{hd_0},
    \label{eq:gdef}
\end{equation}
where $hd_0$ is the final created surface upon full sample rupture. The estimated $G$ is shown for each skin specimen in table\,\ref{table:summary} and is in the $80$ to $210$\,kJ\,m\textsuperscript{-2} range ($135$\,kJ\,m\textsuperscript{-2} in average for all samples, with a significant standard deviation of $35$\,kJ\,m\textsuperscript{-2}).
\\A first remark is that the magnitude of $G$, here reported for the tearing of skin, is significantly higher than that reported for the scissors cutting of skin\,\cite{skinG}, which is about $2\,$kJ\,m\textsuperscript{-2}. It is also higher than the likely energy release rate of individual polymeric fibers (e.g.,\,of collagen fibers, which compose most of the cutaneous tissue), which should be\,\cite{thin_silk} in the order of $1\,$kJ\,m\textsuperscript{-2}. Likely, this difference translates that, contrarily to cutting, tearing is a process involving some fiber-to-fiber interactions rather than only processes below the fiber's scale, with the (likely heat-emitting) friction between these fibers known to account for most of the tissue toughness (e.g.,\,\cite{tearskin}).\\
Another remark is that Vincent-Dospital\,\textit{et al.}\,\cite{TVD2,TVD3} proposed the energy release rate of a material to be related to the core length scale $l$ at which most of the energy is dissipated:
\begin{equation}
    l \sim \frac{a^3G}{2u},
    \label{eq:Gl}
\end{equation}
where $a\sim2$ \AA\,\,is the typical size of a molecular link and $u\sim1\,$eV the typical magnitude of its energetic cohesion. Satisfyingly, in our case, this value is in the micrometer range (l$\sim3\,\upmu$m in average, although Eq.\,(\ref{eq:Gl}) only provides an order a magnitude). Such value is a typical length scale for the diameter of collagen fibers\,\cite{collagen}, which tends to confirm the importance of this scale in the tearing of skin. Note that, because this size is small compared to the extend of our measured thermal anomaly (i.e., Fig.\,\ref{fig:result}), it suggests that most of this anomaly is subsequent to the heat diffusion at larger scale, and not directly related to the intrinsic size of the heat sources $l$.

\subsection{Approximate thermal energy budget\label{sec:budget}}

\begin{figure}[b]
  \includegraphics[width=1\linewidth]{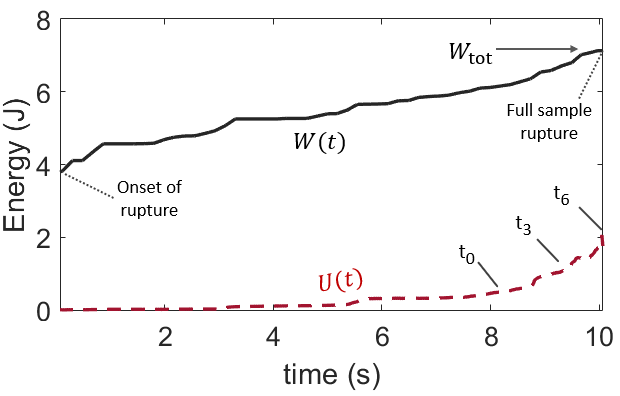}
  \caption{(a): Comparison of the time evolution of the provided mechanical load $W$, as per Fig.\,\ref{fig:force_disp}, and that of the rise in internal (thermal) energy $U$ in the skin sample, as per Eq.\,(\ref{eq:U}). This plot is for skin specimen V. Labels $t_0$ to $t_6$ refer to the frames of Fig.\,\ref{fig:result}. The mean observable thermal efficiency, which is the ratio $\phi=U/W_\text{tot}$ at the end of the rupture, is here about $30$\%.}
  \label{fig:ener}
\end{figure}

To estimate the portion $\phi$ of $W_\text{tot}$ that was dissipated as heat, we computed the rise in internal energy $U$ that could be captured by the infrared camera as
\begin{equation}
    U(t) =  \rho ch\iint_S \Delta T(x,y,t) \,\mathrm{d}s.
    \label{eq:U}
\end{equation}
In this expression, $c$ is the heat capacity of porcine skin, $\sim3.2\,$kJ\,K\textsuperscript{-1}\,kg\textsuperscript{-1} \cite{skin_speheat, skin_speheat2}, $S$ the surface of the sample available to the camera, $x$ and $y$ are the 2D-coordinates of the infrared frames (i.e.,\,see Fig.\,\ref{fig:result}), and $\mathrm{d}s=\mathrm{d}x\times\mathrm{d}y$ is the elementary surface unit (i.e., the surface of one infrared pixel, in this case). We assessed, with a simple scale, the volumetric mass $\rho$ of samples of various measured volumes to be $1150\pm50$\,kg\,m\textsuperscript{-3}.\\
The time evolutions of both the mechanical work $W(t)$ provided to the teared samples (which is defined the same way as $W_\text{tot}$, but for an ongoing rupture) and of the thermal energy $U(t)$ are shown in Fig.\,\ref{fig:ener}, for the same experiment (V) whose results are displayed in Figs.\,\ref{fig:result} and\,\ref{fig:force_disp}. They are also shown for the other skin specimens in appendix\,\ref{app:all}. The mean thermal efficiency $\phi$, for the complete rupture, can then be computed at the end of the experiment, when all of $W_\text{tot}$ as been dissipated. It is defined as the final $U/W_\text{tot}$ ratio and ranges between $5$\% to up to $50$\% depending on the skin sample.
\\Such a wide range for $\phi$ does not come as a surprise. In addition to the likely complexity in the rupture of skin, and to the intrinsic diversity in this biological material, the total dissipated energy and the thermal energy were only broadly estimated. Indeed, our definition of $W_\text{tot}$ (i.e.,\,Eq.\,(\ref{eq:work})) is not fully intrinsic to the studied material and may depend on the loading geometry and sample size. Note that such dependency with the sample size does not however display in table\,\ref{table:summary}, where $l_0$ and $G$ do not appear to be particularly correlated. Likely, it translates that size effects are small compared to our samples' variability in strength. There are also several hypothesis presupposed by Eq.\,(\ref{eq:U}) in the computation of $U$, and not all of them may be conservatively respected. First, it is supposed that all of the thermal energy is available to the infrared camera. In this regard, we made sure to compute $U$ only as long as the heat conduction through the skin did not obviously transfer this energy out of the camera frame (as the full length of the stretched sample is not monitored by the camera which is only framed around the crack tip). We also verified that the energy exchange with the air surrounding the sample was slow enough to be neglected over the time of observation. The typical time constant for such an air-skin exchange was indeed measured (see appendix\,\ref{app:air}) to be about $6$\,minutes when the fracture of a skin sample typically took a few seconds. Likely, the strongest hypothesis behind Eq.\,(\ref{eq:U}) is that the temperature profile, that is only measured at the surface of the epidermis, holds on the full sample's thickness $h$. In practice, skin is a layered (heterogeneous) material, and significant temperature differences may hold between the epidermis, dermis and hypodermis. Additionally, in Fig.\,\ref{fig:result}, one can observe likely thin fiber bundles around the progressing crack, indicating that the assumption that $h$ is a homogeneous thickness is no doubt limited. Finally, as the rupture progresses, the portions of the skin sample lying behind the crack tip gain some freedom in moving outside of the focal plane of the infrared camera, so that the temperature measurement may there be less accurate.\\
Overall, $\phi$ should not be interpreted beyond its order of magnitude. Yet, and despite the listed limitations, we compute, in the next section, the thermal efficiency $\phi$ with a different method and obtain similar results (in order of magnitude) to what we have here reported.

\subsection{Temperature elevation versus damage speed}

\begin{figure}
  \includegraphics[width=1\linewidth]{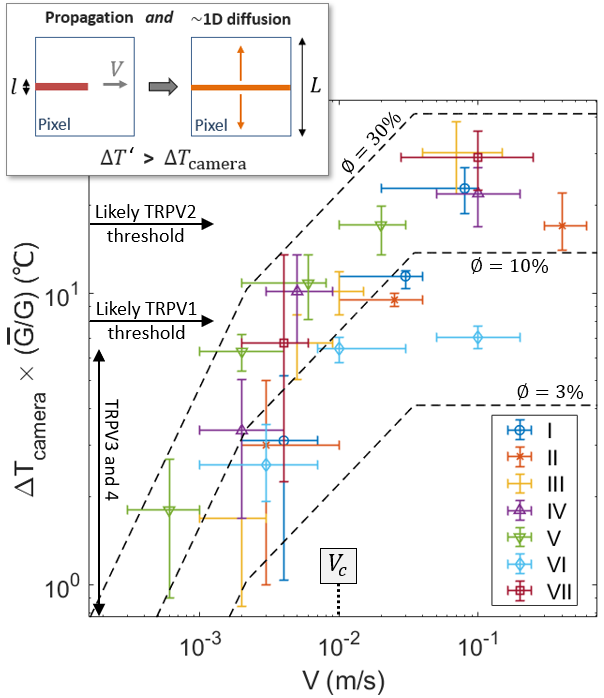}
  \caption{Recorded elevation of temperature $\Delta T$ at the damage tip as a function of crack velocity (log scales), scaled by $\overline{G}/G$ for each skin sample. Data points for the seven experiments are shown with different symbols (and colours). For reference, the thresholds for the activation of the TRPVs are shown by the plain arrows, assuming an ambient temperature of $35^\circ$C for live inner skin. The dashed lines are the model described by Eq.\,(\ref{eq:camera}) using $\phi=3$\%, $10$\% and $30$\%, and with $\rho c =3.7$\,MJ\,K\textsuperscript{-1}\,m\textsuperscript{-3}, $\lambda=0.3$\,J\,s\,\textsuperscript{-1}\,m\,\textsuperscript{-1}K\,\textsuperscript{-1}, $\overline{G}=135$\,kJ\,m\,\textsuperscript{-2} and $l=3\,\upmu$m. The transition velocity $V_c$ predicted by the model (i.e.,\,Eq.\,(\ref{eq:vtrans})) is also indicated. The top-left inset illustrates the 1D diffusion hypothesis lying being Eq.\,(\ref{eq:camera}), due to the time scale difference between the crack propagation and the heat diffusion at the camera pixel size.}
  \label{fig:vel}
\end{figure}

One can notice, in Fig.\,\ref{fig:result}, some correlation between the crack velocity $V$ and the magnitude of the temperature anomaly $\Delta T$. Compare, for instance, the relative crack advancement and the tip temperature between times $t_0$ and $t_1$ and times $t_5$ and $t_6$. Figure\,\ref{fig:vel} displays the relationship between the maximal recorded temperature elevation and the crack velocity, as observed during the experiments. To better compare the different experiments, $\Delta T$ is there rescaled by the ratio $\overline{G}/G$ where $\overline{G}=135\,$kJ\,m\textsuperscript{-2} is the mean energy release rate of all the samples. To avoid confusion, we remind here that $G$ itself is an average value over the rupture of a unique sample (i.e., derived from Eqs.\,(\ref{eq:work}) and\,(\ref{eq:gdef})). 
Note that the exact position of the crack tip, which is necessary to define an accurate velocity and which we have manually picked on each infrared frame, is subject to a large incertitude, and the data in Fig.\,\ref{fig:vel} thus retains relatively large error bars. A similar trend is yet shown for all skin specimens, with $\Delta T$ increasing with the fracture velocity.
\\Such a correlation of temperature elevation with velocity does not come as a surprise, and has been investigated for the rupture of other materials (e.g.,\,\cite{ToussaintSoft}). Fast cracks tend to be hotter, as less time is then allowed for thermal conduction to efficiently evacuate the excess in heat away from the crack tip\,\cite{RiceLevy, ToussaintSoft, TVD2}, where the energy is dissipated. A model based on these considerations has notably shown (see for instance Refs.\,\cite{ToussaintSoft, TVD2}) that, at low velocity, the tip temperature elevation (hereafter referred to as $\Delta T'_\text{slow}(V)$) should increase linearly with fracture velocity. For faster crack tips, the temperature should however increase slower and slower with $V$ (a relation which is hereafter referred to as $\Delta T'_\text{transition}(V)$) and eventually reaches a plateau $\Delta T'_\text{plateau}$ at the highest velocities. The notation $\Delta T'$ is here used to differentiate the prediction of this simple model and the actual camera measured temperature elevation $\Delta T$. The three asymptotic regimes of the model\,\cite{ToussaintSoft, TVD2} are described by:
\begin{equation}
    \Delta T'_\text{slow}\sim \phi G \frac{V}{\lambda},
    \label{eq:slow}
\end{equation}
\begin{equation}
    \Delta T'_\text{transition}\sim \phi G \sqrt{\frac{V}{4\pi \rho c \lambda l}},
    \label{eq:transition}
\end{equation}
\begin{equation}
    \Delta T'_\text{plateau}\sim \frac{\phi G}{\pi \rho c l}.
    \label{eq:plateau}
\end{equation}
where $\lambda \sim 0.3\,$J\,s\,\textsuperscript{-1}\,m\,\textsuperscript{-1}K\,\textsuperscript{-1} is the heat conductivity of skin\,\cite{skin_thermal} and $l$ is the typical length scale over which energy is dissipated and partly transformed into heat.
\\Assuming that the different physical parameters are relatively independent on velocity, these equations describe a transition between $\Delta T' \propto V^{1}$ and $\Delta T' \propto V^{0}$, which is highly compatible with the experimental observation (see Fig.\,\ref{fig:vel}, where $\Delta T$ increasing by $1.5$ orders of magnitude when $V$ increases by $3$ orders of magnitude). A prediction of the model is that such a transition occurs at velocities around $V=V_c$, with
\begin{equation}
V_\text{c}\sim\frac{\lambda}{\pi \rho c l},
\label{eq:vtrans}
\end{equation}
which corresponds\,\cite{ToussaintSoft} to velocities for which the diffusion skin depth $\sim\sqrt{\lambda(l/V)/(\pi \rho c)}$ over the intrinsic warming time $l/V$ is similar to the size $l$ of the heat source. Because our experimental data seems to lie in the transition range, Eq.\,(\ref{eq:vtrans}) is another way of estimating $l$. Indeed, if one uses $V_c\sim1$\,cm\,s\textsuperscript{-1}, as the central order of magnitude of the velocity of our experimental cracks (i.e.,\,see Fig.\,\ref{fig:vel}), one obtains $l$ in the order of a few micrometers. This value is satisfyingly consistent with the prior estimation from Eq.\,(\ref{eq:Gl}) and with the diameter of collagen fibers\,\cite{collagen}.
\\Note that the temperature elevations $\Delta T'$, predicted by the model, only hold at the length scale for heat dissipation $l$. That is, they hold at a scale almost two orders of magnitude smaller than the camera pixel size $L\sim85\,\upmu$m. If one then considers that the heat deposited behind the crack tip diffuses perpendicularly to the crack direction up to the pixel size (i.e.,\,assuming 1D diffusion), the temperature elevation available to our infrared camera, when the extra heat has diffused enough, should then be of the order of:
\begin{equation}
        \Delta T_\text{camera}(V)\sim \frac{l}{L} \times
        \left\{
        \begin{matrix}
        \Delta T'_\text{slow}(V), \hspace{0.8cm}\text{if } V\ll V_c\\
        \Delta T'_\text{transition}(V), \hspace{0.1cm}\text{if } V\sim V_c\\
        \Delta T'_\text{plateau}, \hspace{1.0cm}\text{if } V\gg V_c
        \end{matrix}\right.
    \label{eq:camera}
\end{equation}
This 1D simplification of heat diffusion is here approximately valid because, for the propagation velocities that we consider, the crack advances by one pixel in a time $L/V\sim0.001$\,to $0.1$\,s shorter than the typical time $L^2\rho c/\lambda\sim0.1\,$s for diffusion at the pixel scale. Therefore, at this scale, the heat transport along the crack direction can be approximately neglected (i.e., see the inset of Fig.\,\ref{fig:vel}), in particular for the upper part of our measured fracture velocities.\\
Fitting Eq.\,(\ref{eq:camera}) to the experimental data, as shown in Fig.\,\ref{fig:vel}, one gets a reasonable match, and, then, a new way of estimating $\phi$, that is, a new way of providing an energy budget for the heat dissipation. Indeed, $\phi$ is the only unknown physical quantity in our model. We found $\phi\sim3$ to $30$\%, which is rather compatible with the coarse estimation of section\,\ref{sec:budget}. Again, one should consider this value as only an order of magnitude, as it is, in practice, highly dependent on a model bound to only be a broad representation of the actual tearing of skin (i.e.,\,the estimation of $\phi$ is highly dependent on Eqs.\,(\ref{eq:slow}) to\,(\ref{eq:plateau}) and on the re-scaling proposed by Eq.\,(\ref{eq:camera})). This model is indeed a continuous mesoscopic approach while skin is highly heterogeneous at the fiber's scale, and simple Fourier conduction (a base for the model) is also known to hold limitations to describe the cutaneous heat transport (e.g.,\,\cite{skin_nonfourier}). It nonetheless provides another indication that thermal dissipation accounts for a non negligible part of the strength of the cutaneous tissue.

\section{Conclusion}

In this work, we have demonstrated that the tearing of skin generates temperature anomalies from a few degrees Celsius to tens of degrees Celsius. Unfrozen porcine skin was used as a model for human skin. The recorded heat bursts were observed on space and time scales similar (although slightly coarser) to the expected resolutions of the human neural system\,\cite{neurite_density,TRP_response2,TRP_response}, and in the sensibility range of the thermal biosensor TRPV3, TRPV4 and TRPM2 (often associated to the feeling of warmth\,\cite{PainTRP,skinTRP}), that of TRPV1 and TRPM3 (that have been associated to thermal pain\,\cite{PainTRP,skinTRP}) and that of TRPV2 (which we chose to here include for completeness, although the role of this protein in thermal sensing is uncertain\,\cite{non_pain_TRPV2}). A novel - thermal - pain pathway is thus shown to be likely involved in the reporting of mechanical damages to the nervous system, in particular when the damage rate is fast enough. Indeed, the elevation of skin temperature increases with the damaging rate, and our infrared camera could spot temperature elevations that have the ability to trigger TRPV1 for crack velocities above $1\,$cm\,s\textsuperscript{-1} and to trigger TRPV2 for even faster cracks (see Fig.\,\ref{fig:vel}).\\
In addition to these main results, we have characterised the tearing of our porcine skin samples, by providing their typical mean rupture stress ($\overline{\sigma}_\text{f}\sim12\pm4$\,MPa), their Young modulus ($E\sim20$ to $110$\,MPa depending on the sample), their mean energy release rate ($G\sim135\pm35$\,kJ\,m\textsuperscript{-2}), the heat energy release rate $\phi$ (from a few percent of $G$ to up to $50$\% of $G$, with a most representative value above $10$\%), and the typical length scale for the release of heat ($l$ in the micrometer range as per Eqs.\,(\ref{eq:Gl}) and\,(\ref{eq:vtrans})).\\
We finally showed that a simple physical model\,\cite{ToussaintSoft}, accounting for the heat dissipation and diffusion around cutaneous damages (i.e., Eqs.\,(\ref{eq:slow}) to\,(\ref{eq:camera})), can quantitatively account for the observed dependency of fracture temperature with fracture speed.

\section{Discussion}

\subsection{Direct mechanical algesia and secondary hyperalgesia mechanisms}

From our results, we propose two different thermo-mechanical pain processes. For fastest fractures, direct algesia may arise from the simple activation of TRPV1 (or even TRPV2) at noxious heat level. Slower fractures may not trigger such direct mechanism, although some action potentials may already be send through the nervous system by the warmed-up TRPV3, TRPV4 or TRPM2. It is to be noted that TRPV3 has been shown\,\cite{TRPhysteresis} to present a highly hysteresical sensitivity to thermal anomalies, responding far better to low-intensity temperature bursts after a first activation at a higher, noxious, level. After a first fast damage, very slow ruptures may thus actively be reported via the stronger activation of TRPV3, in a hyperalgesia process. Although it is debated (e.g.\,\cite{non_hyper_TRPV1}), the mitigation of mechanical hyperalgesia, that is, of the increased sensibility to pain after a first stimulus, when suppressing TRVPs, has actually been one of the indications that first suggested that these proteins should play a role in mechanical pain\,\cite{TRPV1_mechano,TRPV1_mechano2,TRPV3_mechano}. Similarly to TRPV3, the TRPV2 channel also holds a strong hysteresis\,\cite{TRP_usedep} and could also, then, play some role in hyperalgesia, while, oppositely, TRPV1 was shown to provide a consistent response to repeated thermal bursts\,\cite{TRP_usedep}. Another, not mutually exclusive, mechanism for hyperalgesia, in our framework, could be that inflamed and/or infected tissues around pre-existing wounds tend to exhibit a higher background temperature (by $1$\,to $5^\circ$C, e.g.,\,\cite{inflinfec}), and could then be sensitive to slower fractures, as the TRPV1 threshold (for instance) shall then be easier to reach. It was also suggested\,\cite{TRPV1_acidification} that existing injuries facilitate the activation of TRPV1 through the acidification of the skin, as this sensor is also responsive to abnormal pH.\\
Interestingly, tissue cooling is already used for anesthesia prior to the mechanical injections of treatments\,\cite{cool_anesthesia, cool_anesthesia2}, and a principle lying being such anesthesia methods may lie in preventing any damage-related thermal anomaly to exceed the nociceptors' thresholds. Additionally, we suggest that using highly conductive materials for the design or the coating of needles and other invasive medical tools (i.e.,\,sharps) may help reduce pain, as the heat may then be efficiently transported away from the tissues through the conductive blades. Actually, most sharps are made of stainless or carbon steel, which are already relatively good thermal conductors but less so than some other metals or some other specific materials.

\subsection{On possible higher temperatures at the smallest scales}

The pixel size of our infrared measurement ($\sim85\,\upmu$m) was about half an order of magnitude bigger than the typical distance between two neurites ($\sim20\,\upmu$m\,\cite{neurite_density, TVDpain1}). It was also almost two orders of magnitude bigger than the length scale $l$, where the heat is dissipated (here inverted to be in the micrometer range, which is similar to the size of the skin fibers, with two different methods, i.e., with Eqs.\,(\ref{eq:Gl}) or\,(\ref{eq:vtrans})). We therefore suggest that, contrarily to what is suggested by Fig.\,\ref{fig:vel}, cracks propagating at slower velocities than $1\,$cm\,s\textsuperscript{-1} may already be locally hot enough to trigger direct algesia. Indeed, and although this was not here measured, high local thermal anomalies (above the TRPV1 threshold) may exist at the neurites or fibers' scale for these lower velocities. In the most extreme scenario, Eq.\,(\ref{eq:plateau}) predicts temperature burst up to $\Delta T'\sim\phi G/(\pi Cl)\sim100 - 1000^\circ$C at the microscopic scale. Of course such temperatures may seem excessive in a biological context, but they were actually suspected in the rupture of various other materials\,\cite{ToussaintSoft, Bouchaud2012, Fuller1975, TVD2}. These high temperatures could themselves cause a rapid deterioration of the surrounding skin cells (e.g.,\,\cite{thermal_biodamage}), but they would only occur briefly and locally around an already failing tissue so that such potential secondary thermal damage may not be of key significance.\\
What remains certain is that milder temperature anomalies, but still strong enough to trigger the TRP nociceptors, can be directly measured.

\subsection{In vivo versus ex vivo skin}

Although porcine skin is only a model of human skin, we have discussed, in section\,\ref{sec:sample}, how the orders of magnitude we report should be valid to the human biology, in particular as pig and human skin have relatively close mechanical strength and thermal properties\,\cite{PigHuman, PigHuman2, skin_speheat}.\\
An additional complexity to consider, in order to assess the relevance of our results to live subjects, is that live skin is continuously blood-irrigated, which is a source for heat transport (advection) not at play in our experiments. In the skin capillaries, which are likely present locally around any rupture, an order of magnitude for the blood velocity at rest is $V_B \sim 0.5\,$mm\,s\,\textsuperscript{-1} (e.g.,\,\cite{blood_velocity}). Even considering a perfect and instantaneous heat transfer between the structural tissue and such a blood network (a hardly realistic hypothesis), this would imply a smear of the dissipated heat over, at most, a distance $\tau V_B \sim 10\,\upmu$m, for the time interval $\tau=20\,$ms accessible to our camera. This non-conservative length scale is in the same range and actually slightly smaller than the skin depth of the simultaneous heat conduction in our samples $\sqrt{\lambda\tau/(\rho c)}\sim40\,\upmu$m. Thus, we point out that our experimental lack of blood circulation shall not significantly undermine our core observations. For similar reasons, it is also unlikely that a hormonal regulation of the temperature, also bound to occur in an in vivo skin, can efficiently mitigate the transient heat signal we describe, as hormonal transport (either in the blood or by diffusion in the tissue (e.g.,\,\cite{paracine})) should be slow compared to the generation of a skin fracture.\\
Note however that, to some extent, further studies may target in vivo human samples (rather than ex vivo porcine ones), as skin rupture is a common medical practice and as infrared measurements are fully non-invasive.

\subsection{On the magnitude of the heat dissipation and various damage types}

In this manuscript, we gave a broad estimation, with two different methods, of the percentage of mechanical work that is effectively transformed into heat during the propagation of cutaneous cracks ($\phi\sim3$ to $50$\%, with a mean representative value above $10$\% - see table\,\ref{table:summary} and Fig.\,\ref{fig:vel}). This portion being non negligible, it could make thermal monitoring a ´natural' way for the detection of damages, and, as a very qualitative statement, it would not be surprising that evolution exploited the detection of the dissipated heat for the preservation of life. Note that, consistently, our estimation of $\phi$ is inline with what we elsewhere reported for the tear of another fibrous tissue of biological origin, that is, paper\,\cite{ToussaintSoft}, where $\phi$ was measured between $10$\% and $40$\%.
\\The here measured thermal anomalies are however bigger than those that we recently theorised, when the discussed pain pathway was first proposed\,\cite{TVDpain1}. In this former theoretical study, it was suggested that these anomalies should be on the edge of the TRPs sensitivity and thus relevant to hyperalgesia only rather than to direct algesia as well. By contrast, we have here shown that direct thermo-mechanical algesia is also likely at play for fast cracks. One of the differences between the previous theoretical work and the present one is that damages at the full skin scale have here been studied, while the rupture of a unique collagen fiber was priorly considered. This being stated, the main difference lies in the value of the considered energy release rate, that was here computed to be in the $80-210$\,kJ\,m\textsuperscript{-2} range for the tearing of skin, but reported to be more than one order of magnitude smaller for the cutting of skin\,\cite{skinG}. Such discrepancy likely derives from the different role of inter fiber friction when tearing or cutting skin. Such inner friction was proposed to account for most of the cutaneous strength in tearing\,\cite{tearskin}, but is likely negligible in cutting. Note that the tearing of porcine dermis, with different loading modes than the one that we here studied, can also dissipate less energy than our inverted mean energy release rate\,\cite{Persson_skin}. Similar studies to the present one should then be performed for other types of damages, and in particular for cuts or punctures which are both common injuries and common medical procedures. Thermo-mechanical pain may there be of different importance.

\subsection{On other pain mechanisms}

It is important to state that the pain pathway that we here propose is not to comprehensively account for any sense of pain. For instance, the pressure pain threshold in human subjects was measured (e.g.,\,\cite{PainPressThreh}) to be around $0.1$\,to $1$\,MPa, that is, at stress levels far less than what is needed to initiate an actual skin rupture and hence strong thermal anomalies (i.e., $\overline{\sigma}_\text{f}$ in the order of $10\,$MPa, as here or elsewhere\,\cite{other_young} reported). It is actually comforting that pain shall occur before an actual rupture, but shall also increase, maybe through different mechanisms, when rupture is indeed reached.\\
Other nociceptors exist at the membrane of neurons, for instance, the Piezo channels\,\cite{piezo2_nox}, which opening is believed to be related to the stretch of cells' membranes. Such channel opening with stretch has, interestingly, also been proposed as another explanation for the involvement of TRPs in mechanical sensing\,\cite{forcing_trps}, without the consideration of any thermal anomaly. In practice, both effects could coexist, with the thermal sensitivity of TRPs that could be improved by their abnormal stretch, hence leading to a polymodal detection.\\
Finally, in this work, we have only considered the triggering of heat sensitive TRPs as the mechanism for mechanical pain. One may additionally wonder about a similar involvement of the cold sensitive TRPs, such as TRPA1 and TRPM8, as an effect of these proteins on mechanical algesia has also been evidenced\,(e.g.,\,\cite{TRPA1_mech, TRPM8_mech}). Some cross-talk has been shown to take place between TRPs, for instance between TRPV1 and TRPA1\,\cite{crosstalk_TRP}, and a hypothesis could be that the abnormal activation of the former by fracture-induced heat may alter the responsiveness of the latter. In our experiments, we have not registered any local decrease of skin temperature around the tearing fractures and only measured an temperature increase, but such cold anomalies - which could directly trigger cold sensitive TRPs - have actually been reported in the rupture of other materials\,\cite{Fuller1975,Rittel_1998}. Additionally, in the cutaneous tissue, the vasoconstriction under mechanical stress could also be prone to induce some local cooling\,\cite{blood_temperature}.
\\Overall, mechanical algesia is bound to be a very convoluted phenomenon, involving many types of nociceptors and of biological processes\,\cite{mechano_pain}. The present study aimed to introduce and shed some (infrared) light on a new, thermo-mechanical, pain pathway.\\

\section*{Contributions, Acknowledgements, conflicts of interest and ethics}

\noindent
T.V.-D. proposed the pain theory and performed the experimental work, R.T. and K.J.M. advised on rupture mechanics and energy dissipation and on the experimental set-up. T.V.-D. wrote the first version of the manuscript and all authors contributed and agreed on the final version.\\
The authors acknowledge the support of the University of Oslo, of PoreLab, of the Njord centre and of the IRP France-Norway D-FFRACT. This work was also partly supported by the Research Council of Norway through its Centre of Excellence funding scheme,  project number 262644. The authors declare no competing financial interests in the publishing of this manuscript.\\
We thank Zbigniew Rozynek from the Adam Mickiewicz University for its early experimental assistance and for insightful discussions. We also thank the Strøm-Larsen butchery for providing the porcine skin and answering rather unusual questions about it. The skin was a standard product of the shop, so that no animal was harmed specifically for the purpose of the present work.

\newpage
\appendix

\section{Other skin specimens\label{app:all}}

\begin{figure*}
  \includegraphics[width=1\linewidth]{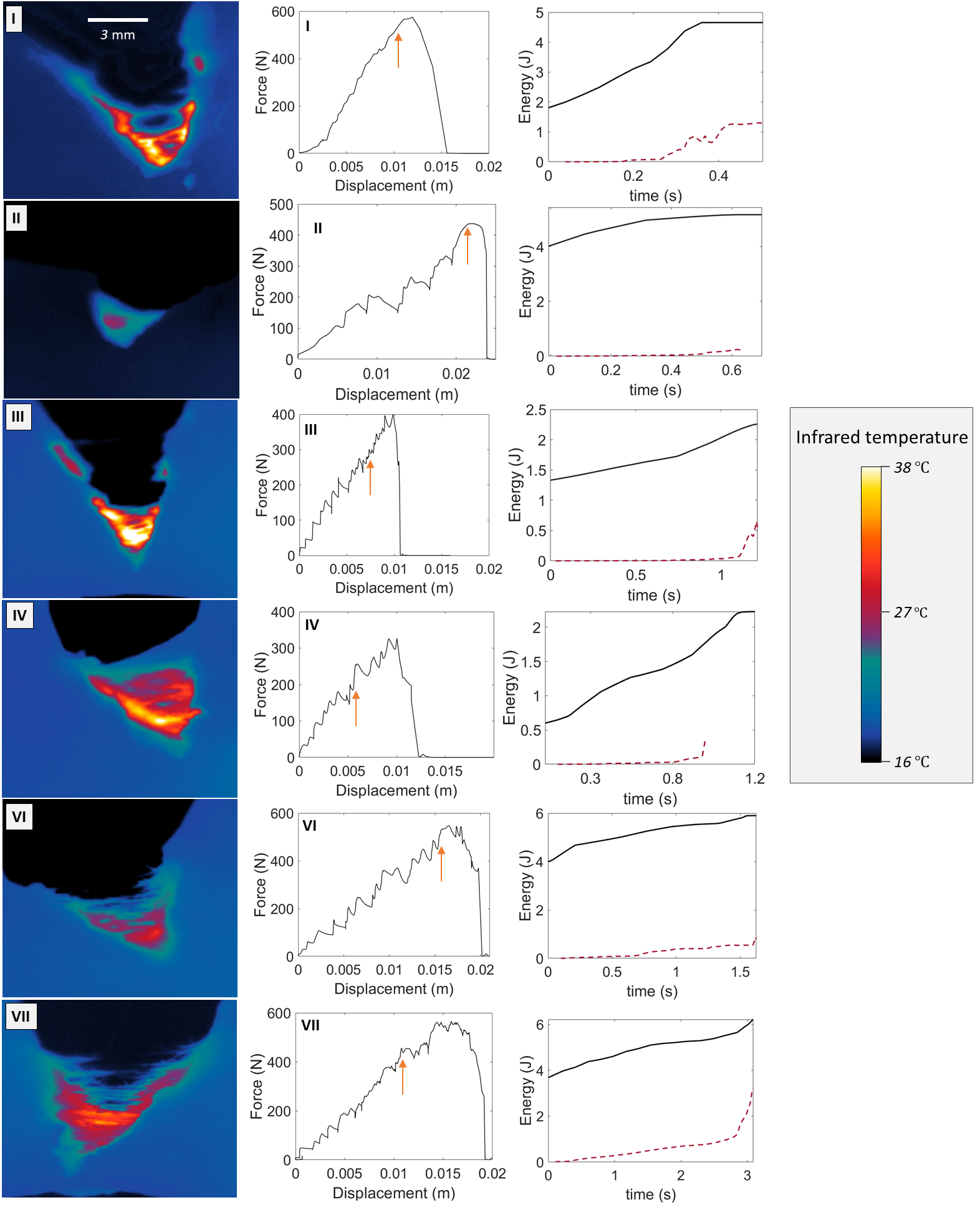}
  \caption{Temperature map examples (first column), force versus displacement curves (second column) and time evolution of $W$ and $U$ (plain and dashed plot respectively, third column) from the onset of the rupture (arbitrarily at $t=0$\,s) for skin specimens I, II, III, IV, VI and VII. The arrows on the $F$ versus $\Delta l$ plots indicate the onsets of rupture.}
  \label{fig:result2}
\end{figure*}

Figure\,\ref{fig:result2} shows the experimental measurements for the skin specimens not fully presented in the manuscript (i.e., all specimens except sample V). A temperature map at an arbitrary time, the force versus displacement plot and the time evolution of $W$ and $U$ after the onset of rupture are there represented.

\section{Thermal exchange with the ambient air\label{app:air}}

\begin{figure}
  \includegraphics[width=1\linewidth]{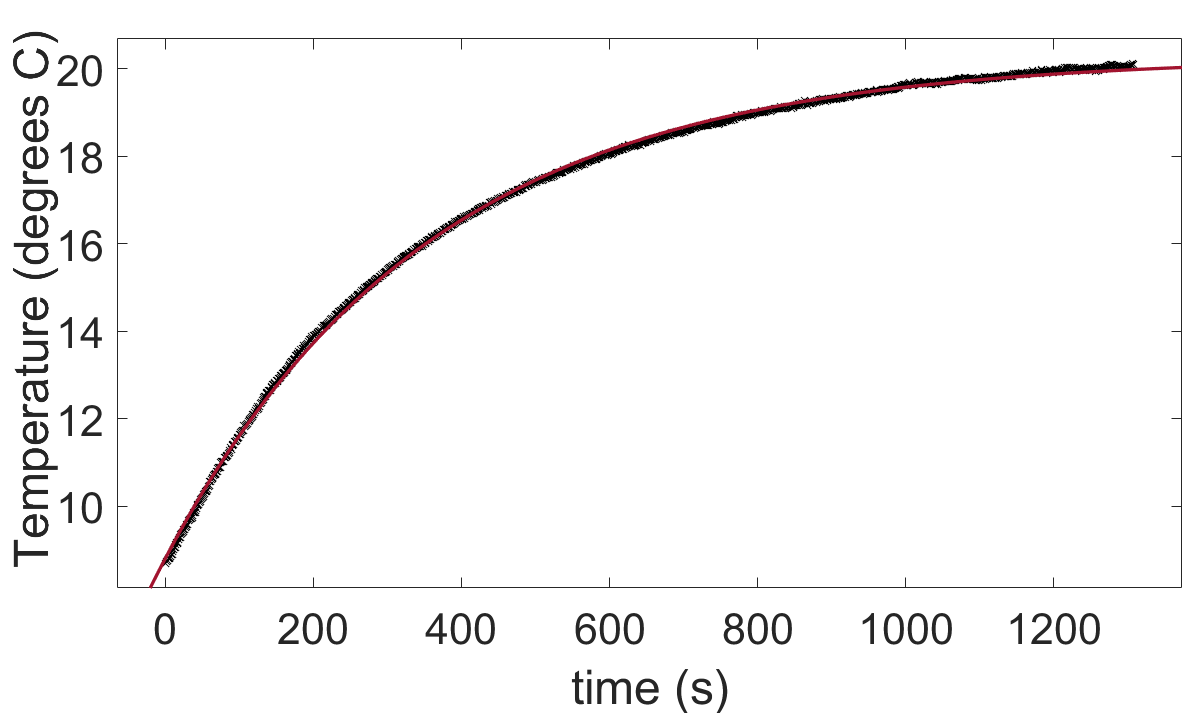}
  \caption{Warming up in air at room temperature of a skin sample taken out from a refrigerator. The data points (crosses) are measured with the infrared camera. The plain line is a fit of Eq.\,(\ref{eq:exchange}) with a time constant $\tau_\text{a}=356$\,s.}
  \label{fig:warming}
\end{figure}

Figure\,\ref{fig:warming} shows the slow warming up of a skin sample taken out of a refrigerator and placed, in the ambient air, in front of the infrared camera. The following function is fitted to the temperature data points:
\begin{equation}
    T(t) =  T_\text{i}+\Delta T_\text{max}\left(1-\exp\left(-\,\cfrac{t}{\tau_\text{a}}\right)\right),
    \label{eq:exchange}
\end{equation}
where $T_\text{i}$ is the initial sample temperature, $\Delta T_\text{max}$ is the final rise in temperature of the warning sample and $\tau_\text{a}\sim6$\,min is the typical time constant for the air-skin thermal exchange. This time constant being about two orders of magnitude higher than the typical duration of our fracture tests (from $1$\,to $10\,$s), the air-skin thermal exchange has been neglected in this manuscript.


\FloatBarrier

\begin{thebibliography}{74}
\providecommand{\natexlab}[1]{#1}
\providecommand{\url}[1]{\texttt{#1}}
\expandafter\ifx\csname urlstyle\endcsname\relax
  \providecommand{\doi}[1]{doi: #1}\else
  \providecommand{\doi}{doi: \begingroup \urlstyle{rm}\Url}\fi

\bibitem[Griffith(1921)]{Griffith1921}
A.~Griffith.
\newblock {The Phenomena of Rupture and Flow in Solids}.
\newblock \emph{Philosophical Transactions of the Royal Society of London A:
  Mathematical, Physical and Engineering Sciences}, 221\penalty0
  (582-593):\penalty0 163--198, January 1921.
\newblock ISSN 1471-2962.
\newblock \doi{10.1098/rsta.1921.0006}.

\bibitem[Rice and Drucker(1967)]{rice_surface}
J.~R. Rice and D.~C. Drucker.
\newblock {Energy changes in stressed bodies due to void and crack growth}.
\newblock \emph{International Journal of Fracture Mechanics}, 3:\penalty0
  19--27, 1967.
\newblock ISSN 1573-2673.
\newblock \doi{10.1007/BF00188642}.

\bibitem[Morrissey and Rice(1998)]{crackwaves}
J.~W. Morrissey and J.~R. Rice.
\newblock Crack front waves.
\newblock \emph{Journal of the Mechanics and Physics of Solids}, 46\penalty0
  (3):\penalty0 467 -- 487, 1998.
\newblock ISSN 0022-5096.
\newblock \doi{10.1016/S0022-5096(97)00072-0}.

\bibitem[Irwin(1957)]{Irwin1957}
G.~R. Irwin.
\newblock {Analysis of stresses and strains near the end of a crack traversing
  a plate}.
\newblock \emph{Journal of Applied Mechanics}, 24:\penalty0 361--364, 1957.

\bibitem[Rice and Levy(1969)]{RiceLevy}
J.~R. Rice and N.~Levy.
\newblock Local heating by plastic deformation at a crack tip.
\newblock \emph{Physics of Strength and Plasticity}, pages 277--293, 1969.

\bibitem[Fuller et~al.(1975)Fuller, Fox, and Field]{Fuller1975}
K.~N.~G. Fuller, P.~G. Fox, and J.~E. Field.
\newblock The temperature rise at the tip of fast-moving cracks in glassy
  polymers.
\newblock \emph{Proceedings of the Royal Society of London A: Mathematical,
  Physical and Engineering Sciences}, 341\penalty0 (1627):\penalty0 537--557,
  1975.
\newblock ISSN 0080-4630.
\newblock \doi{10.1098/rspa.1975.0007}.

\bibitem[Pallares et~al.(2012)Pallares, Rountree, Douillard, Charra, and
  Bouchaud]{Bouchaud2012}
G.~Pallares, C.~L. Rountree, L.~Douillard, F.~Charra, and E.~Bouchaud.
\newblock Fractoluminescence characterization of the energy dissipated during
  fast fracture of glass.
\newblock \emph{Europhysics Letters}, 99\penalty0 (2):\penalty0 28003, 2012.

\bibitem[Vincent-Dospital et~al.(2020{\natexlab{a}})Vincent-Dospital,
  Toussaint, Santucci, Vanel, Bonamy, Hattali, Cochard, Flekkøy, and
  Måløy]{TVD2}
T.~Vincent-Dospital, R.~Toussaint, S.~Santucci, L.~Vanel, D.~Bonamy,
  L.~Hattali, A.~Cochard, E.~G. Flekkøy, and K.~J. Måløy.
\newblock How heat controls fracture: the thermodynamics of creeping and
  avalanching cracks.
\newblock \emph{Soft Matter}, 16:\penalty0 9590--9602, 2020{\natexlab{a}}.
\newblock \doi{10.1039/D0SM01062F}.

\bibitem[Toussaint et~al.(2016)Toussaint, Lengliné, Santucci,
  Vincent-Dospital, Naert-Guillot, and Måløy]{ToussaintSoft}
R.~Toussaint, O.~Lengliné, S.~Santucci, T.~Vincent-Dospital, M.~Naert-Guillot,
  and K.~J. Måløy.
\newblock How cracks are hot and cool: a burning issue for paper.
\newblock \emph{Soft Matter}, 12:\penalty0 5563--5571, 2016.
\newblock \doi{10.1039/C6SM00615A}.

\bibitem[Marshall et~al.(1974)Marshall, Coutts, and Williams]{Marshall_1974}
G.~P. Marshall, L.~H. Coutts, and J.~G. Williams.
\newblock Temperature effects in the fracture of {PMMA}.
\newblock \emph{Journal of Materials Science}, 9\penalty0 (9):\penalty0
  1409--1419, Sep 1974.
\newblock ISSN 1573-4803.
\newblock \doi{10.1007/BF00552926}.

\bibitem[Carbone and Persson(2005)]{carbonePersson}
G.~Carbone and B.~N.~J. Persson.
\newblock Hot cracks in rubber: Origin of the giant toughness of rubberlike
  materials.
\newblock \emph{Phys. Rev. Lett.}, 95:\penalty0 114301, Sep 2005.
\newblock \doi{10.1103/PhysRevLett.95.114301}.

\bibitem[Braeck and Podladchikov(2007)]{ThermalRunaway}
S.~Braeck and Y.~Y. Podladchikov.
\newblock Spontaneous thermal runaway as an ultimate failure mechanism of
  materials.
\newblock \emph{Phys. Rev. Lett.}, 98:\penalty0 095504, Mar 2007.
\newblock \doi{10.1103/PhysRevLett.98.095504}.

\bibitem[Vincent-Dospital et~al.(2020{\natexlab{b}})Vincent-Dospital,
  Toussaint, Cochard, Måløy, and Flekkøy]{TVD1}
T.~Vincent-Dospital, R.~Toussaint, A.~Cochard, K.~J. Måløy, and E.~G.
  Flekkøy.
\newblock Thermal weakening of cracks and brittle-ductile transition of matter:
  {A} phase model.
\newblock \emph{Physical Review Materials}, 02 2020{\natexlab{b}}.
\newblock \doi{10.1103/PhysRevMaterials.4.023604}.

\bibitem[Rice(2006)]{HeatWeak}
J.~R. Rice.
\newblock Heating and weakening of faults during earthquake slip.
\newblock \emph{Journal of Geophysical Research: Solid Earth}, 111\penalty0
  (B5), 2006.
\newblock \doi{10.1029/2005JB004006}.

\bibitem[Wibberley and Shimamoto(2005)]{pressur2005}
C.~Wibberley and T.~Shimamoto.
\newblock Earthquake slip weakening and asperities explained by fluid
  pressurization.
\newblock \emph{Nature}, 436:\penalty0 689--92, 09 2005.

\bibitem[Sulem and Famin(2009)]{SulemCarbo}
J.~Sulem and V.~Famin.
\newblock Thermal decomposition of carbonates in fault zones: Slip-weakening
  and temperature-limiting effects.
\newblock \emph{Journal of Geophysical Research: Solid Earth}, 114\penalty0
  (B3), 2009.
\newblock \doi{10.1029/2008JB006004}.

\bibitem[Vincent-Dospital and Toussaint(2021)]{TVDpain1}
T.~Vincent-Dospital and R.~Toussaint.
\newblock Thermo-mechanical pain: the signaling role of heat dissipation in
  biological tissues.
\newblock \emph{New Journal of Physics}, 23\penalty0 (2):\penalty0 023028, feb
  2021.
\newblock \doi{10.1088/1367-2630/abe444}.

\bibitem[Wang and Woolf(2005)]{PainTRP}
H.~Wang and C.~J. Woolf.
\newblock Pain {TRPs}.
\newblock \emph{Neuron}, 46\penalty0 (1):\penalty0 9 -- 12, 2005.
\newblock ISSN 0896-6273.
\newblock \doi{10.1016/j.neuron.2005.03.011}.

\bibitem[Tóth et~al.(2014)Tóth, Oláh, Szöllősi, and Bíró]{skinTRP}
B.~I. Tóth, A.~Oláh, A.~Gábor Szöllősi, and T.~Bíró.
\newblock {TRP} channels in the skin.
\newblock \emph{British journal of pharmacology}, 171:\penalty0 2568 -- 2581,
  2014.
\newblock \doi{10.1113/jphysiol.2005.088377}.

\bibitem[Nilius et~al.(2005)Nilius, Talavera, Owsianik, Prenen, Droogmans, and
  Voets]{voltage_gating}
B.~Nilius, K.~Talavera, G.~Owsianik, J.~Prenen, G.~Droogmans, and T.~Voets.
\newblock Gating of {TRP} channels: a voltage connection.
\newblock \emph{The Journal of physiology}, 567:\penalty0 35 -- 44, 2005.
\newblock \doi{10.1113/jphysiol.2005.088377}.

\bibitem[Xu et~al.(2002)Xu, Ramsey, Kotecha, Moran, Chong, Lawson, Ge, Lilly,
  Silos-Santiago, Xie, DiStefano, Curtis, and Clapham]{TRVP3_grad2}
H.~Xu, I.~S. Ramsey, S.~A. Kotecha, M.~M. Moran, J.~A. Chong, D.~Lawson, P.~Ge,
  J.~Lilly, I.~Silos-Santiago, Y.~Xie, P.~S. DiStefano, R.~Curtis, and D.~E.
  Clapham.
\newblock {TRPV3} is a calcium-permeable temperature-sensitive cation channel.
\newblock \emph{Nature}, 418:\penalty0 181--186, 2002.
\newblock \doi{10.1038/nature00882}.

\bibitem[Caterina et~al.(2000)Caterina, Leffler, Malmberg, Martin, Trafton,
  Petersen-Zeitz, Koltzenburg, Basbaum, and D.]{TRPV1_julius}
M.J. Caterina, A.~Leffler, A.B. Malmberg, W.J. Martin, J.~Trafton, K.R.
  Petersen-Zeitz, M.~Koltzenburg, A.I. Basbaum, and Julius D.
\newblock Impaired nociception and pain sensation in mice lacking the capsaicin
  receptor.
\newblock \emph{Science}, 288:\penalty0 306 -- 3013, 2000.
\newblock \doi{10.1126/science.288.5464.306}.

\bibitem[G{\"u}ler et~al.(2002)G{\"u}ler, Lee, Iida, Shimizu, Tominaga, and
  Caterina]{TRPV4_temp}
A.~D. G{\"u}ler, H.~Lee, T.~Iida, I.~Shimizu, M.~Tominaga, and M.~Caterina.
\newblock Heat-evoked activation of the ion channel, {TRPV4}.
\newblock \emph{Journal of Neuroscience}, 22\penalty0 (15):\penalty0
  6408--6414, 2002.
\newblock ISSN 0270-6474.
\newblock \doi{10.1523/JNEUROSCI.22-15-06408.2002}.

\bibitem[Kashio and Tominaga(2017)]{TRPM2_temp}
M.~Kashio and M.~Tominaga.
\newblock The {TRPM2} channel: A thermo-sensitive metabolic sensor.
\newblock \emph{Channels}, 11\penalty0 (5):\penalty0 426--433, 2017.
\newblock \doi{10.1080/19336950.2017.1344801}.

\bibitem[Vriens et~al.(2011)Vriens, Owsianik, Hofmann, Philipp, Stab, Chen,
  Benoit, Xue, Janssens, Kerselaers, Oberwinkler, Vennekens, Gudermann, Nilius,
  and Voets]{TRPM3_temp}
J.~Vriens, G.~Owsianik, T.~Hofmann, S.E. Philipp, J.~Stab, X.~Chen, M.~Benoit,
  F.~Xue, A.~Janssens, S.~Kerselaers, J.~Oberwinkler, R.~Vennekens,
  T.~Gudermann, B.~Nilius, and T.~Voets.
\newblock {TRPM3} is a nociceptor channel involved in the detection of noxious
  heat.
\newblock \emph{Neuron}, 70\penalty0 (3):\penalty0 482--494, 2011.
\newblock ISSN 0896-6273.
\newblock \doi{10.1016/j.neuron.2011.02.051}.

\bibitem[Park et~al.(2011)Park, Vastani, Guan, Raja, Koltzenburg, and
  Caterina]{non_pain_TRPV2}
U.~Park, N.~Vastani, Y.~Guan, S.~N. Raja, M.~Koltzenburg, and M.~J. Caterina.
\newblock {TRP V}anilloid 2 knock-out mice are susceptible to perinatal
  lethality but display normal thermal and mechanical nociception.
\newblock \emph{Journal of Neuroscience}, 31\penalty0 (32):\penalty0
  11425--11436, 2011.
\newblock ISSN 0270-6474.
\newblock \doi{10.1523/JNEUROSCI.1384-09.2011}.

\bibitem[Culp et~al.(1989)Culp, Ochoa, Cline, and Dotson]{coupled_pain}
W.~J. Culp, J.~Ochoa, M.~Cline, and R.~Dotson.
\newblock {Heat and mechanical hyperalgesia induced by capsaisin: cross
  modality threshold modulation in human {C} nociceptors}.
\newblock \emph{Brain}, 112\penalty0 (5):\penalty0 1317--1331, 10 1989.
\newblock ISSN 0006-8950.
\newblock \doi{10.1093/brain/112.5.1317}.

\bibitem[Smith(2010)]{cool_anesthesia}
K.~C. Smith.
\newblock Ice anesthesia for injection of dermal fillers.
\newblock \emph{Dermatologic Surgery}, 36:\penalty0 812--814, 2010.
\newblock \doi{10.1111/j.1524-4725.2010.01549.x}.

\bibitem[Besirli et~al.(2020)Besirli, Smith, Zacks, Gardner, Pipe, Musch, and
  Shah]{cool_anesthesia2}
C.~G. Besirli, S.~J. Smith, D.~N. Zacks, T.~W. Gardner, K.~P. Pipe, D.~C.
  Musch, and A.~R. Shah.
\newblock Randomized safety and feasibility trial of ultra-rapid cooling
  anesthesia for intravitreal injections.
\newblock \emph{Ophthalmology Retina}, 4\penalty0 (10):\penalty0 979--986,
  2020.
\newblock ISSN 2468-6530.
\newblock \doi{10.1016/j.oret.2020.04.001}.

\bibitem[Walker et~al.(2003)Walker, Urban, Medhurst, Patel, Panesar, Fox, and
  McIntyre]{TRPV1_mechano}
K.~M. Walker, L.~Urban, S.~J. Medhurst, S.~Patel, M.~Panesar, A.~J. Fox, and
  P.~McIntyre.
\newblock The vr1 antagonist capsazepine reverses mechanical hyperalgesia in
  models of inflammatory and neuropathic pain.
\newblock \emph{Journal of Pharmacology and Experimental Therapeutics},
  304\penalty0 (1):\penalty0 56--62, 2003.
\newblock ISSN 0022-3565.
\newblock \doi{10.1124/jpet.102.042010}.

\bibitem[Pomonis et~al.(2003)Pomonis, Harrison, Mark, Bristol, Valenzano, and
  Walker]{TRPV1_mechano2}
J.~D. Pomonis, J.~E. Harrison, L.~Mark, D.~R. Bristol, K.~J. Valenzano, and
  K.~Walker.
\newblock A novel, orally effective vanilloid receptor 1 antagonist with
  analgesic properties. in vivo characterization in rat models of inflammatory
  and neuropathic pain.
\newblock \emph{Journal of Pharmacology and Experimental Therapeutics},
  306\penalty0 (1):\penalty0 387--393, 2003.
\newblock ISSN 0022-3565.
\newblock \doi{10.1124/jpet.102.046268}.

\bibitem[McGaraughty et~al.(2017)McGaraughty, Chu, Xu, Leys, Radek, Dart,
  Gomtsyan, Schmidt, Kym, and Brederson]{TRPV3_mechano}
S.~McGaraughty, K.~L. Chu, J.~Xu, L.~Leys, R.~J. Radek, M.~J. Dart,
  A.~Gomtsyan, R.~G. Schmidt, P.~R. Kym, and J.-D. Brederson.
\newblock {TRPV3} modulates nociceptive signaling through peripheral and
  supraspinal sites in rats.
\newblock \emph{Journal of neurophysiology}, 118:\penalty0 904--916, 2017.
\newblock ISSN 1522-1598.
\newblock \doi{10.1152/jn.00104.2017}.

\bibitem[Urano et~al.(2012)Urano, Ara, Fujinami, and Hiraoka]{non_hyper_TRPV1}
H.~Urano, T.~Ara, Y.~Fujinami, and B.~Y. Hiraoka.
\newblock Aberrant {TRPV1} expression in heat hyperalgesia associated with
  trigeminal neuropathic pain.
\newblock \emph{International Journal of Medical Sciences}, 9:\penalty0
  690--697, 2012.
\newblock \doi{10.7150/ijms.4706}.

\bibitem[Liedtke et~al.(2003)Liedtke, Tobin, Bargmann, and
  Friedman]{TRPV4_mechano}
Wolfgang Liedtke, David~M. Tobin, Cornelia~I. Bargmann, and Jeffrey~M.
  Friedman.
\newblock Mammalian {TRPV4 (VR-OAC)} directs behavioral responses to osmotic
  and mechanical stimuli in caenorhabditis elegans.
\newblock \emph{Proceedings of the National Academy of Sciences}, 100\penalty0
  (suppl 2):\penalty0 14531--14536, 2003.
\newblock ISSN 0027-8424.
\newblock \doi{10.1073/pnas.2235619100}.

\bibitem[Oaklander(2001)]{neurite_density}
A.-L. Oaklander.
\newblock The density of remaining nerve endings in human skin with and without
  postherpetic neuralgia after shingles.
\newblock \emph{Pain}, 92, 2001.
\newblock ISSN 0304-3959.

\bibitem[Yao et~al.(2010)Yao, Liu, and Qin]{TRP_response2}
J.~Yao, B.~Liu, and F.~Qin.
\newblock Kinetic and energetic analysis of thermally activated {TRPV1}
  channels.
\newblock \emph{Biophysical journal}, 99:\penalty0 1743--1753, 2010.
\newblock ISSN 1542-0086.
\newblock \doi{10.1016/j.bpj.2010.07.022}.

\bibitem[Liu and Qin(2019)]{TRP_response}
B.~Liu and F.~Qin.
\newblock Patch-clamp combined with fast temperature jumps to study thermal
  {TRP} channels.
\newblock In \emph{{TRP} Channels}, volume 1987 of \emph{Methods in Molecular
  Biology}. Humana, New York, NY, 2019.
\newblock \doi{10.1007/978-1-4939-9446-5_9}.

\bibitem[Debeer et~al.(2013)Debeer, Le~Luduec, Kaiserlian, Laurent, Nicolas,
  Dubois, and Kanitakis]{PigHuman}
S.~Debeer, J.B. Le~Luduec, D.~Kaiserlian, P.~Laurent, J.F. Nicolas, B.~Dubois,
  and J.~Kanitakis.
\newblock Comparative histology and immunohistochemistry of porcine versus
  human skin.
\newblock \emph{European Journal of Dermatology}, 23:\penalty0 456--466, 2013.
\newblock ISSN 1167-1122.
\newblock \doi{10.1684/ejd.2013.2060}.

\bibitem[Ranamukhaarachchi et~al.(2016)Ranamukhaarachchi, Lehnert,
  Ranamukhaarachchi, Sprenger, Schneider, Mansoor, Rai, Häfeli, and
  Stoeber]{PigHuman2}
S.~A. Ranamukhaarachchi, S.~Lehnert, S.~L. Ranamukhaarachchi, L.~Sprenger,
  T.~Schneider, I.~Mansoor, K.~Rai, U.~O. Häfeli, and B.~Stoeber.
\newblock A micromechanical comparison of human and porcine skin before and
  after preservation by freezing for medical device development.
\newblock \emph{Scientific Reports}, 6:\penalty0 32074, 2016.
\newblock ISSN 2045-2322.
\newblock \doi{10.1038/srep32074}.

\bibitem[Thomsen et~al.(2014)Thomsen, H.-G., Mathiesen, Poulsen, Sørensen,
  Tarnow, and Feidenhans'l]{skin_mod_examp}
M.~Thomsen, Anier H.-G., J.~Mathiesen, M.~Poulsen, D.~N. Sørensen, L.~Tarnow,
  and R.~Feidenhans'l.
\newblock Model study of the pressure build-up during subcutaneous injection.
\newblock \emph{PLOS ONE}, 9\penalty0 (8):\penalty0 1--7, 08 2014.
\newblock \doi{10.1371/journal.pone.0104054}.

\bibitem[{Otsuka} et~al.(2002){Otsuka}, {Okada}, {Hassan}, and
  {Togawa}]{skin_surf_temp}
K.~{Otsuka}, S.~{Okada}, M.~{Hassan}, and T.~{Togawa}.
\newblock Imaging of skin thermal properties with estimation of ambient
  radiation temperature.
\newblock \emph{IEEE Engineering in Medicine and Biology Magazine}, 21\penalty0
  (6):\penalty0 49--55, 2002.
\newblock \doi{10.1109/MEMB.2002.1175138}.

\bibitem[Saxena and Arya(1981)]{skin_temp1}
V.P. Saxena and D.~Arya.
\newblock Steady-state heat distribution in epidermis, dermis and subdermal
  tissues.
\newblock \emph{Journal of Theoretical Biology}, 89\penalty0 (3):\penalty0 423
  -- 432, 1981.
\newblock ISSN 0022-5193.

\bibitem[Henriques and Moritz(1947)]{skin_speheat2}
F.~C. Henriques and A.~R. Moritz.
\newblock Studies of thermal injury: {I.} the conduction of heat to and through
  skin and the temperatures attained therein. a theoretical and an experimental
  investigation.
\newblock \emph{The American journal of pathology}, 23:\penalty0 530 -- 549,
  1947.

\bibitem[Cohen(1977)]{skin_thermal}
M.~L. Cohen.
\newblock Measurement of the thermal properties of human skin. a review.
\newblock \emph{Journal of Investigative Dermatology}, 69\penalty0
  (3):\penalty0 333 -- 338, 1977.
\newblock ISSN 0022-202X.
\newblock \doi{10.1111/1523-1747.ep12507965}.

\bibitem[Scanga(2005)]{meat_processing}
J.A. Scanga.
\newblock Slaughter and fabrication/boning processes and procedures.
\newblock In \emph{Improving the Safety of Fresh Meat}, Woodhead Publishing
  Series in Food Science, Technology and Nutrition, pages 259--272. Woodhead
  Publishing, 2005.
\newblock ISBN 978-1-85573-955-0.
\newblock \doi{10.1533/9781845691028.2.259}.

\bibitem[Leonard et~al.(2021)Leonard, Worden, Boettcher, Dickinson, and
  Hartstone-Rose]{freezing}
K.~C. Leonard, N.~Worden, M.~L. Boettcher, E.~Dickinson, and A.~Hartstone-Rose.
\newblock Effects of freezing and short-term fixation on muscle mass, volume,
  and density.
\newblock \emph{The Anatomical Record}, 2021.
\newblock \doi{10.1002/ar.24639}.

\bibitem[Jain and Sharma(1998)]{blackbody}
P~Jain and L.~Sharma.
\newblock The physics of blackbody radiation: A review.
\newblock \emph{Journal of Applied Science in Southern Africa}, 4:\penalty0
  80--101, 02 1998.
\newblock \doi{10.4314/jassa.v4i2.16899}.

\bibitem[Soerensen et~al.(2014)Soerensen, Clausen, Mercer, and
  Pedersen]{pig_emissivity}
D.~D. Soerensen, S.~Clausen, J.~B. Mercer, and L.~J. Pedersen.
\newblock Determining the emissivity of pig skin for accurate infrared
  thermography.
\newblock \emph{Computers and Electronics in Agriculture}, 109:\penalty0
  52--58, 2014.
\newblock ISSN 0168-1699.
\newblock \doi{10.1016/j.compag.2014.09.003}.

\bibitem[Miyazaki and Hayashi(1999)]{single_fiber}
H.~Miyazaki and K.~Hayashi.
\newblock Tensile tests of collagen fibers obtained from the rabbit patellar
  tendon.
\newblock \emph{Biomedical Microdevices}, 2:\penalty0 151--157, 1999.
\newblock ISSN 1572-8781.
\newblock \doi{10.1023/A:1009953805658}.

\bibitem[{Ní Annaidh} et~al.(2012){Ní Annaidh}, Bruyère, Destrade,
  Gilchrist, and Otténio]{other_young}
A.~{Ní Annaidh}, K.~Bruyère, M.~Destrade, M.~D. Gilchrist, and M.~Otténio.
\newblock Characterization of the anisotropic mechanical properties of excised
  human skin.
\newblock \emph{Journal of the Mechanical Behavior of Biomedical Materials},
  5\penalty0 (1):\penalty0 139--148, 2012.
\newblock ISSN 1751-6161.
\newblock \doi{10.1016/j.jmbbm.2011.08.016}.

\bibitem[Rodriguez et~al.(2021)Rodriguez, Mangiagalli, and
  Persson]{Persson_skin}
N.~Rodriguez, P.~Mangiagalli, and B.~N.~J. Persson.
\newblock \emph{Viscoelastic Crack Propagation: Review of Theories and
  Applications}, pages 377--420.
\newblock Springer International Publishing, 2021.
\newblock ISBN 978-3-030-68920-9.
\newblock \doi{10.1007/12_2020_76}.

\bibitem[Pereira et~al.(1997)Pereira, Lucas, and Swee-Hin]{skinG}
B.~P. Pereira, P.~W. Lucas, and T.~Swee-Hin.
\newblock Ranking the fracture toughness of thin mammalian soft tissues using
  the scissors cutting test.
\newblock \emph{Journal of Biomechanics}, 30\penalty0 (1):\penalty0 91 -- 94,
  1997.
\newblock ISSN 0021-9290.
\newblock \doi{10.1016/S0021-9290(96)00101-7}.

\bibitem[Porter et~al.(2013)Porter, Guan, and Vollrath]{thin_silk}
D.~Porter, J.~Guan, and F.~Vollrath.
\newblock Spider silk: Super material or thin fibre?
\newblock \emph{Advanced Materials}, 25\penalty0 (9):\penalty0 1275--1279,
  2013.
\newblock \doi{10.1002/adma.201204158}.

\bibitem[Yang et~al.(2015)Yang, Sherman, Gludovatz, Schaible, Stewart, Ritchie,
  and Meyers]{tearskin}
W.~Yang, V.~R. Sherman, B.~Gludovatz, E.~Schaible, P.~Stewart, R.~O. Ritchie,
  and M.~A. Meyers.
\newblock On the tear resistance of skin.
\newblock \emph{Nature Communications}, 6:\penalty0 6649, 2015.
\newblock ISSN 2041-1723.
\newblock \doi{10.1038/ncomms7649}.

\bibitem[Vincent-Dospital et~al.(2021)Vincent-Dospital, Toussaint, Cochard,
  Flekkøy, and Måløy]{TVD3}
T.~Vincent-Dospital, R.~Toussaint, A.~Cochard, E.~G. Flekkøy, and K.~J.
  Måløy.
\newblock Thermal dissipation as both the strength and weakness of matter. a
  material failure prediction by monitoring creep.
\newblock \emph{Soft Matter}, 2021.
\newblock \doi{10.1039/D0SM02089C}.

\bibitem[Verhaegen et~al.(2012)Verhaegen, Van~Marle, Kuehne, Schouten, Gaffney,
  Maini, Middelkoop, and Van~Zuijlen]{collagen}
P.~D.H.M. Verhaegen, J.~Van~Marle, A.~Kuehne, H.~J. Schouten, E.~A. Gaffney,
  P.~K. Maini, E.~Middelkoop, and P.~P.M. Van~Zuijlen.
\newblock Collagen bundle morphometry in skin and scar tissue: a novel distance
  mapping method provides superior measurements compared to fourier analysis.
\newblock \emph{Journal of Microscopy}, 245\penalty0 (1):\penalty0 82--89,
  2012.
\newblock \doi{10.1111/j.1365-2818.2011.03547.x}.

\bibitem[Giering et~al.(1996)Giering, Lamprecht, and Minet]{skin_speheat}
K.~Giering, I.~Lamprecht, and O.~Minet.
\newblock {Specific heat capacities of human and animal tissues}.
\newblock In \emph{Laser-Tissue Interaction and Tissue Optics}, pages 188 --
  197. International Society for Optics and Photonics, SPIE, 1996.
\newblock \doi{10.1117/12.229547}.

\bibitem[Hooshmand et~al.(2015)Hooshmand, Moradi, and Khezry]{skin_nonfourier}
P.~Hooshmand, A.~Moradi, and B.~Khezry.
\newblock Bioheat transfer analysis of biological tissues induced by laser
  irradiation.
\newblock \emph{International Journal of Thermal Sciences}, 90:\penalty0 214 --
  223, 2015.
\newblock ISSN 1290-0729.
\newblock \doi{10.1016/j.ijthermalsci.2014.12.004}.

\bibitem[Liu et~al.(2011)Liu, Yao, Zhu, and Qin]{TRPhysteresis}
B.~Liu, J.~Yao, M.~X. Zhu, and F.~Qin.
\newblock {Hysteresis of gating underlines sensitization of {TRPV3} channels}.
\newblock \emph{Journal of General Physiology}, 138\penalty0 (5):\penalty0
  509--520, 10 2011.
\newblock ISSN 0022-1295.
\newblock \doi{10.1085/jgp.201110689}.

\bibitem[Liu and Qin(2016)]{TRP_usedep}
B.~Liu and F.~Qin.
\newblock Use dependence of heat sensitivity of vanilloid receptor {TRPV2}.
\newblock \emph{Biophysical journal}, 110:\penalty0 1523--1537, 2016.
\newblock ISSN 1542-0086.
\newblock \doi{10.1016/j.bpj.2016.03.005}.

\bibitem[Chanmugam et~al.(2017)Chanmugam, Langemo, Thomason, Haan, Altenburger,
  Tippett, and Zortman]{inflinfec}
A.~Chanmugam, D.~Langemo, K.~Thomason, J.~Haan, E.A. Altenburger, L.~Tippett,
  A.~Henderson, and T.A. Zortman.
\newblock Relative temperature maximum in wound infection and inflammation as
  compared with a control subject using long-wave infrared thermography.
\newblock \emph{Adv Skin Wound Care}, 30:\penalty0 406--414, 2017.
\newblock \doi{10.1097/01.ASW.0000522161.13573.62}.

\bibitem[Tominaga et~al.(1998)Tominaga, Caterina, Malmberg, Rosen, Gilbert,
  Skinner, Raumann, Basbaum, and Julius]{TRPV1_acidification}
M.~Tominaga, M.~J. Caterina, A.~B. Malmberg, T.~A. Rosen, H.~Gilbert,
  K.~Skinner, B.~E. Raumann, A.~I. Basbaum, and D.~Julius.
\newblock The cloned capsaicin receptor integrates multiple pain-producing
  stimuli.
\newblock \emph{Neuron}, 21\penalty0 (3):\penalty0 531--543, 1998.
\newblock ISSN 0896-6273.
\newblock \doi{10.1016/S0896-6273(00)80564-4}.

\bibitem[Xu and Lu(2011)]{thermal_biodamage}
F.~Xu and T.~Lu.
\newblock \emph{Skin Bioheat Transfer and Skin Thermal Damage}, pages 23--68.
\newblock Springer Berlin Heidelberg, 2011.
\newblock ISBN 978-3-642-13202-5.
\newblock \doi{10.1007/978-3-642-13202-5_3}.

\bibitem[Stücker et~al.(1996)Stücker, Baier, Reuther, Hoffmann, Kellam, and
  Altmeyer]{blood_velocity}
M.~Stücker, V.~Baier, T.~Reuther, K.~Hoffmann, K.~Kellam, and P.~Altmeyer.
\newblock Capillary blood cell velocity in human skin capillaries located
  perpendicularly to the skin surface: Measured by a new laser doppler
  anemometer.
\newblock \emph{Microvascular Research}, 52\penalty0 (2):\penalty0 188--192,
  1996.
\newblock ISSN 0026-2862.
\newblock \doi{10.1006/mvre.1996.0054}.

\bibitem[Berry et~al.(2015)Berry, Godara, Liovic, and Haylock]{paracine}
J.~D. Berry, P.~Godara, P.~Liovic, and D.~N. Haylock.
\newblock Predictions for optimal mitigation of paracrine inhibitory signalling
  in haemopoietic stem cell cultures.
\newblock \emph{Stem Cell Research \& Therapy}, 6:\penalty0 58, 2015.
\newblock ISSN 1757-6512.
\newblock \doi{10.1186/s13287-015-0048-7}.

\bibitem[Jensen et~al.(1986)Jensen, Andersen, Olesen, and
  Lindblom]{PainPressThreh}
K.~Jensen, H.~Ø. Andersen, J.~Olesen, and U.~Lindblom.
\newblock Pressure-pain threshold in human temporal region. evaluation of a new
  pressure algometer.
\newblock \emph{Pain}, 25\penalty0 (3):\penalty0 313--323, 1986.
\newblock ISSN 0304-3959.
\newblock \doi{10.1016/0304-3959(86)90235-6}.

\bibitem[Murthy et~al.(2018)Murthy, Loud, Daou, Marshall, Schwaller,
  K{\"u}hnemund, Francisco, Keenan, Dubin, Lewin, and Patapoutian]{piezo2_nox}
S.~E. Murthy, M.~C. Loud, I.~Daou, K.~L. Marshall, F.~Schwaller,
  J.~K{\"u}hnemund, A.~G. Francisco, W.~T. Keenan, A.~E. Dubin, G.~R. Lewin,
  and A.~Patapoutian.
\newblock The mechanosensitive ion channel {Piezo2} mediates sensitivity to
  mechanical pain in mice.
\newblock \emph{Science Translational Medicine}, 10\penalty0 (462), 2018.
\newblock ISSN 1946-6234.
\newblock \doi{10.1126/scitranslmed.aat9897}.

\bibitem[Liu and Montell(2015)]{forcing_trps}
C.~Liu and C.~Montell.
\newblock Forcing open {TRP} channels: Mechanical gating as a unifying
  activation mechanism.
\newblock \emph{Biochemical and Biophysical Research Communications},
  460\penalty0 (1):\penalty0 22 -- 25, 2015.
\newblock ISSN 0006-291X.
\newblock \doi{10.1016/j.bbrc.2015.02.067}.

\bibitem[Brierley et~al.(2011)Brierley, Castro, Harrington, Hughes, Page,
  Rychkov, and Blackshaw]{TRPA1_mech}
S.~M. Brierley, J.~Castro, A.~M. Harrington, P.~A. Hughes, A.~J. Page, G.~Y.
  Rychkov, and L.~A. Blackshaw.
\newblock {TRPA1} contributes to specific mechanically activated currents and
  sensory neuron mechanical hypersensitivity.
\newblock \emph{The Journal of physiology}, 589:\penalty0 3575--3593, 2011.
\newblock \doi{10.1113/jphysiol.2011.206789}.

\bibitem[De~Caro et~al.(2018)De~Caro, Russo, Avagliano, Cristiano, Calignano,
  Aramini, Bianchini, Allegretti, and Brandolini]{TRPM8_mech}
C.~De~Caro, R.~Russo, C.~Avagliano, C.~Cristiano, A.~Calignano, A.~Aramini,
  G.~Bianchini, M.~Allegretti, and L.~Brandolini.
\newblock Antinociceptive effect of two novel transient receptor potential
  melastatin 8 antagonists in acute and chronic pain models in rat.
\newblock \emph{British Journal of Pharmacology}, 175\penalty0 (10):\penalty0
  1691--1706, 2018.
\newblock \doi{10.1111/bph.14177}.

\bibitem[Spahn et~al.(2014)Spahn, Stein, and Z{\"o}llner]{crosstalk_TRP}
V.~Spahn, C.~Stein, and C.~Z{\"o}llner.
\newblock Modulation of transient receptor vanilloid 1 activity by transient
  receptor potential ankyrin 1.
\newblock \emph{Molecular Pharmacology}, 85\penalty0 (2):\penalty0 335--344,
  2014.
\newblock ISSN 0026-895X.
\newblock \doi{10.1124/mol.113.088997}.

\bibitem[Rittel(1998)]{Rittel_1998}
D.~Rittel.
\newblock Experimental investigation of transient thermoelastic effects in
  dynamic fracture.
\newblock \emph{International Journal of Solids and Structures}, 35\penalty0
  (22):\penalty0 2959--2973, 1998.
\newblock ISSN 0020-7683.
\newblock \doi{10.1016/S0020-7683(97)00352-1}.

\bibitem[Rubinstein and Sessler(1990)]{blood_temperature}
E.-H. Rubinstein and Daniel I. Sessler.
\newblock {Skin-surface Temperature Gradients Correlate with Fingertip Blood
  Flow in Humans}.
\newblock \emph{Anesthesiology: The Journal of the American Society of
  Anesthesiologists}, 73\penalty0 (3):\penalty0 541--545, 09 1990.
\newblock ISSN 0003-3022.

\bibitem[Hill and Bautista(2020)]{mechano_pain}
R.~Z. Hill and D.~M. Bautista.
\newblock Getting in touch with mechanical pain mechanisms.
\newblock \emph{Trends in Neurosciences}, 43:\penalty0 311 -- 325, 2020.
\newblock ISSN 0166-2236.
\newblock \doi{10.1016/j.tins.2020.03.004}.

\end{thebibliography}


\end{document}